\documentclass[12pt,a4paper]{article}

\usepackage{amsmath,amssymb,color,graphics,graphicx}
\usepackage{bm}
\usepackage{hyperref} 
\RequirePackage{graphicx}

\definecolor{dark-green}{rgb}{0,0.7,0}
\definecolor{dark-blue}{rgb}{0,0.2,0.5}
\definecolor{med-blue}{rgb}{0,0.7,1}
\definecolor{mblue}{rgb}{0,0.2,1}
\definecolor{cnc}{rgb}{0.8,0,0}
\definecolor{light-red}{rgb}{1,0.8,0.8}
\definecolor{dark-yellow}{rgb}{1,0.8,0}
\definecolor{light-blue}{rgb}{0.8,0.9,1}
\definecolor{verylight-blue}{rgb}{0.93,0.95,1}
\definecolor{light-yellow}{rgb}{1,0.9,0.8}
\definecolor{grey}{gray}{0.88}

\def\a{\alpha}
\def\b{\beta}
\def\c{\gamma}

\def\d{\delta}
\def\g{\gamma}

\def\vt{\vartheta}

\def\stareq{\stackrel{*}{=}}

\usepackage{amsfonts}

\begin{document}

\title{Conservation of energy-momentum of matter as the basis for
  the gauge theory of gravitation}

\author{ Friedrich W.\ Hehl$^{1,\star}$ and Yuri N. Obukhov$^{2,\ast}$\\
  \small $^1$Institute of Theoretical Physics, University of Cologne, 50923 K\"oln, Germany\\
  \small $^2$Institute for Nuclear Safety, Russian Academy of Sciences, 115191 Moscow, Russia\\
  \small $^{\star}$hehl@thp.uni-koeln.de\qquad $^{\ast}$obukhov@ibrae.ac.ru}

\maketitle

\begin{abstract}
According to Yang \& Mills (1954), a {\it conserved} current and a related rigid (`global') symmetry lie at the foundations of gauge theory. When the rigid symmetry is extended to a {\it local} one, a so-called gauge symmetry, a new interaction emerges as gauge potential $A$; its field strength is $F\sim {\rm curl} A$. In gravity, the conservation of the energy-momentum current of matter and the rigid translation symmetry in the Minkowski space of special relativity lie at the foundations of a gravitational gauge theory. If the translation invariance is made local, a gravitational potential $\vt$ arises together with its field strength $T\sim {\rm curl}\,\vt$. Thereby the Minkowski space deforms into a Weitzenb\"ock space with nonvanishing torsion $T$ but vanishing curvature. The corresponding theory is reviewed and its equivalence to general relativity pointed out. Since translations form a subgroup of the Poincar\'e group, the group of motion of special relativity, one ought to straightforwardly extend the gauging of the translations to the gauging of full Poincar\'e group thereby also including the conservation law of the {\it angular momentum} current. The emerging Poincar\'e gauge (theory of) gravity, starting from the viable Einstein-Cartan theory of 1961, will be shortly reviewed and its prospects for further developments assessed.
\end{abstract}
\newpage

{\hypersetup{linkcolor=black}
\tableofcontents}

\section{Yang-Mills theory, gauge theory}

In the 1920s and 1930s it became clear that the atomic nuclei consist of protons ($p$) and neutrons ($n$) which interact with each other via a {\it strong} nuclear force. The masses of proton and neutron are nearly equal. The proton carries a positive elementary electric charge whereas the neutron is electrically neutral (but still caries a magnetic moment). Otherwise, in particular with respect to their nuclear interaction, they behave very similar. This charge independence of the nuclear interaction of $p$-$p$, $n$-$p$, and $n$-$n$ was an important {\it experimental} result.

Heisenberg \cite{Heisenberg:1932} was led to the hypothesis that there exists a new particle called {\it nucleon} that has two different {\it states}, a positively charged one, the proton, and a neutral one, the neutron. These two different states were put in analogy to an electron which can have a state with spin up and one with spin down. Accordingly, Heisenberg attributed to the nucleon the new quantum number $\bm I$ of isospin, which is {\it conserved} in nuclear interactions. And the isospin up, $I_3=+\frac 12$, represents the proton and the one down $I_3=- \frac 12$, the neutron.
 
After Yukawa \cite{Yukawa:1935} had introduced the pion $\pi$ as mediator of the strong nuclear force, it eventually turned out that the pion exists in three differently charged states, namely as $\pi^{+}$, $\pi^-$, and as $\pi^0$. Thus, one had to attribute to it the isospin $I=1$. With the help of this insight, one got a consistent and experimentally verified framework for the nuclear force. At the same time, the new quantum number isospin found its way from nuclear physics into the systematics of elementary particle physics, as proposed by Kemmer \cite{Kemmer:1939}.

Considering the nucleon together with the pion, it became clear that the invariance group of the strong nuclear interaction at the level of the nucleon is the unitary Lie group $SU(2)$ and the charge independence of the nuclear interaction translates into the requirement that no direction in the isospin space is distinguished. In other words, the corresponding action is invariant under rigid $SU(2)$ transformations and we have an associated {\it conservation} of the isospin ${\bm I}$.

Here Yang \& Mills (1954) set in, proposing ``Conservation of Isotopic Spin and Isotopic Gauge Invariance'' as the foundation for establishing a hypothetical $SU(2)$ gauge theory of {\it strong interaction} \cite{Yang:1954}. The conserved isospin current, via the reciprocal of the Noether theorem \cite{Kosmann}, yields a rigid (`global') $SU(2)$-invariance. Insisting, as Yang and Mills did, that a rigid symmetry is inconsistent with field-theoretical ideas, the $SU(2)$-invariance is postulated to be valid {\it locally.} This enforces to introduce a compensating (or gauge) field $A$, the gauge potential,\footnote{Yang
\& Mills denoted it with $B$ in their original paper \cite{Yang:1954}.} which upholds the $SU(2)$-invariance even under these generalized local transformations. Then the curl of $A$ turns out to be the field strength of the emerging gauge field. The prototypical procedure for the conserved electric current of the Dirac Lagrangian and its $U(1)$ gauge invariance had already been executed by Weyl \cite{Weyl:1929} and Fock \cite{Fock:1929} in 1929, see also \cite{Weyl:1950}. Accordingly, we can define a gauge theory as follows:

  {\it A gauge theory is a heuristic scheme within the Lagrange
    formalism in the Minkowski space of special relativity for the
    purpose of deriving a new interaction from a conserved current
    and the attached rigid symmetry group. This new
    `gauge' interaction is induced by demanding that the rigid
    symmetry should be extended to a locally valid symmetry.}

Explicitly, the Yang-Mills type gauging works as follows, see, e.g., O'Rai\-fertaigh \cite{ORaifeartaigh:1978jea}, Mack \cite{Mack:1979es}, or Chaichian \& Nelipa \cite{Chaichian:1984}. Let $I = \int d^4x\,{\mathfrak L}$ be an action for the matter field $\psi^A$ with the Lagrangian density ${\mathfrak L} = {\mathfrak L}(\psi^A, \partial_i\psi^A)$. We transform the matter field under the rigid action of an $N$-parameter internal symmetry group $G$ 
\begin{equation}
\psi^A \longrightarrow \psi'^A = \psi^A + \delta\psi^A,\qquad
\delta\psi^A = \varepsilon^I(t_I)^A_B\,\psi^B,\label{delP}
\end{equation}
with the generators $t_I$, $I = 1, \dots, N$, and $\partial_i\varepsilon^I = 0$. We suppose that the action does not change under the transformation of the matter field: $\delta I = 0$. We assume $G$ to be a Lie group, and the generators $t_I\in {\cal G}$ form the basis of the corresponding Lie algebra with the commutator
\begin{equation}
\left[t_I\,,t_J\right] = f^K{}_{IJ}\,t_K.\label{comm}
\end{equation}
The structure constants $f^K{}_{IJ} = -\,f^K{}_{JI}$ satisfy the Jacobi identity 
\begin{equation}
f^N{}_{IL}f^L{}_{JK} + f^N{}_{JL}f^L{}_{KI} + f^N{}_{KL}f^L{}_{IJ} \equiv 0.\label{Jacobi} 
\end{equation}

The {\it Noether theorem} tells us that, provided the matter variables satisfy the field
equations, the invariance of the action under (\ref{delP}) yields a conservation law
\begin{equation}
\delta I = 0\qquad \Longrightarrow\qquad \partial_i J^i{}_I = 0\label{consJ}
\end{equation}
of the canonical {\it Noether current}
\begin{equation}\label{J}
J^i{}_I := (t_I)^A_B\,\psi^B\,{\frac {\partial {\mathfrak L}}{\partial \partial_i\psi^A}}.
\end{equation}
As a result, for an $N$-parameter symmetry group there exist $N$ conserved charges
\begin{equation}
Q_I = \int d^3x\,J^0{}_I,\qquad I = 1,\dots, N,\label{chargeN}
\end{equation}
where integral is taken over the spatial 3-surface $t =\,$const. 

When the symmetry is made local, $\partial_i\varepsilon^I \neq 0$, the action with the matter Lagrangian
${\mathfrak L}(\psi^A, \partial_i\psi^A)$ is no longer invariant. One needs a gauge (compensating) field $A_i{}^I$ to be introduced via the {\it minimal coupling} recipe
\begin{equation}
  {\mathfrak L}(\psi^A, \partial_i\psi^A) \longrightarrow {\mathfrak L}(\psi^A, D_i\psi^A),
\end{equation}
with the partial derivative replaced $\partial_i\rightarrow D_i$ by the {\it covariant} one:
\begin{equation}
D_i\psi^A = \partial_i\psi^A + A_i{}^I\,(t_I)^A_B\,\psi^B.\label{Dpsi}
\end{equation}
Then the invariance of the modified action $I = \int d^4x\,{\mathfrak L}(\psi^A, D_i\psi^A)$ is recovered because the crucial covariance property
\begin{equation}
\delta (D_i\psi^A) = \varepsilon^I(x)\,(t_I)^A_B\,D_i\psi^B\label{delDP}
\end{equation}
is guaranteed by the inhomogeneous transformation law of the gauge field:
\begin{equation}\label{delA}
  \delta A_i{}^I = -\,D_i\varepsilon^I = -\,(\partial_i\varepsilon^I + A_i{}^Kf^I{}_{KJ}\varepsilon^J).
\end{equation}

This completes the kinematics of the gauge theory. The gauge field $A_i{}^I$ becomes a true dynamical variable by adding a suitable kinetic term, ${\mathfrak V}$, to the minimally coupled matter Lagrangian: ${\mathfrak L}\rightarrow{\mathfrak L} + {\mathfrak V}$. This supplementary term has to be gauge invariant, such that the gauge invariance of the total action is kept.  The gauge invariance of ${\mathfrak V}$ is obtained by constructing it in terms of the {\it gauge field strength}:
\begin{equation}
  F_{ij}{}^I = \partial_i A_j{}^I - \partial_j A_i{}^I + f^I{}_{JK}A_i{}^J A_j{}^K.\label{FYM}
\end{equation}
Using (\ref{delA}) and (\ref{Jacobi}) we straightforwardly verify the transformation law $\delta F_{ij}{}^I = \varepsilon^K(x)\,f^I{}_{KJ}F_{ij}{}^J$. The important property of the gauge field strength is the {\it Bianchi identity}
\begin{equation}
D_{[k}F_{ij]}{}^I = 0,\label{BiaF}
\end{equation}
which can be naturally interpreted as the homogeneous field equation. 

Since the gauge field Lagrangian ${\mathfrak V}$ should be also invariant under the local symmetry group, it should be a function of $F_{ij}{}^I$. The (inhomogeneous) Yang--Mills field equation is derived from the total action
\begin{equation}\label{Itot}
  I_{\rm tot} = \int d^4x\left\{{\mathfrak L}(\psi^A, D_i\psi^A)
    + {\mathfrak V}(F_{ij}{}^I)\right\}.
\end{equation}
Variation with respect to the gauge field potential yields explicitly
\begin{equation}
  D_j H^{ij}{}_I = J^i{}_I,\qquad\text{with}\qquad
  H^{ij}{}_I := -\,2{\frac {\partial {\mathfrak V}}{\partial F_{ij}{}^I}}.\label{DHJ}
\end{equation}
Quite remarkably, the matter source of the gauge field turns out to be a covariant Noether current (\ref{J}). However, in the locally gauge invariant theory, the original conservation law (\ref{consJ}) is replaced by the covariant one
\begin{equation}
D_i J^i{}_I = 0.\label{DJ}
\end{equation}
By recasting (\ref{DHJ}) into
\begin{equation}
  \partial_j H^{ij}{}_I = {\stackrel AJ}{\,}^i{}_I,\qquad
  {\stackrel AJ}{\,}^i{}_I = J^i{}_I  + A_j{}^K f^J{}_{KI}H^{ij}{}_J,\label{DHA}
\end{equation}
we can derive the modified conservation law
\begin{equation}
\partial_i {\stackrel AJ}{\,}^i{}_I = 0,\label{DJA}
\end{equation}
which reflects the fact that the gauge field couples not only to matter, but also to itself.  In other words, the {\it gauge field carries its own charge}. 

As we see, the formal structure of a general gauge field theory (\ref{DJ}), (\ref{BiaF}), and (\ref{DHJ}) appears as a generalization of the Maxwell theory \cite{Birkbook}. The final ``building block'' of this generalization is the explicit form of the {\it constitutive relation} $H = H(F)$ between the gauge field strength and the excitation, see above in \eqref{DHJ}:
\begin{equation}
  H^{ij}{}_I = -\,2{\frac {\partial {\mathfrak V}}{\partial F_{ij}{}^I}}.\label{HijA}
\end{equation}
In the original Yang-Mills theory \cite{Yang:1954}, the Lagrangian was constructed as a quadratic Maxwell type invariant of the gauge field strength, and the resulting constitutive law is linear: $H^{ij}{}_I = F^{ij}{}_I$. Later, Mills \cite{Mills:1979} also discussed a {\it non\/}linear, Born--Infeld type ``constitutive'' relation between $H$ and $F$. But this didn't prove to be useful.

\begin{figure}
\centering
\includegraphics[width=0.55\textwidth]{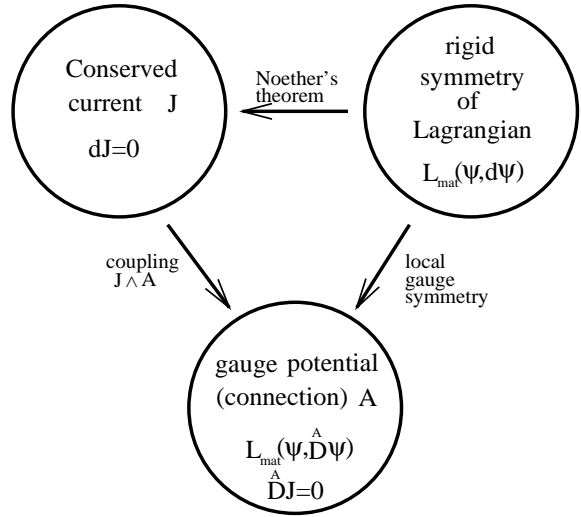}
\caption{
The structure of a gauge theory \`a la Yang--Mills is depicted in this diagram, which is adapted from Mills \cite{Mills:1989}.}\label{YMF}
\end{figure}

Schematically, we represented the gauge procedure in Figure~\ref{YMF}.
\medskip

Let us stress our main points: A gauge theory is based on a {\it conserved current} and the {\it symmetry} connected with it. The symmetry is first rigid---and there is no interaction---then, subsequently, made local, and the gauge potential $A$ and the gauge field strength $F\sim \stackrel{A}{D}\hspace{-1pt}A$ emerge in this procedure.

Incidentally, here we confine our attention only to classical field theory and we do {\it not} investigate quantum field theoretical consequences \cite{Kiefer}. However, a consistent particle picture does arise in the quasiclassical approximation. A point particle with the 4-velocity $u^i$ carries internal degrees of freedom in the form of a ``color charge'' $q_I$.  It couples with the gauge field via the
interaction Lagrangian $q_Iu^iA_i{}^I\!,$ see Wong \cite{Wong:1970}. Accordingly, particle's motion is affected by the generalized Lorentz force
\begin{equation}
f_i = q_I F_{ij}{}^I u^j.\label{LFYM}
\end{equation}

\section{Newton-Einstein gravity}

Turning now to gravity, the fundamental question is: What is the conserved current and what is the gauge group of gravity? Our starting point will be, of course, Newtonian gravity. There mass is the source of gravity or rather the mass density $\rho({\bm r},t)$ in its quasi field-theoretical formulation as Poisson equation for the gravitational potential $\phi({\bm r},t)$:
\begin{equation}\label{Poisson}
  \Delta \phi({\bm r},t)=4\pi G\rho({\bm r},t)\,.
\end{equation}
Here $\Delta$ is the Laplace operator and $G$ Newton's gravitational constant. In Newtonian mechanics, the motion of a material continuum with the mass density $\rho$ is described by the velocity vector field $\bm{v}$. The mass inside the volume $\Omega(t)$ is given by the integral
\begin{equation}
  m(t) = \int\limits_{\Omega(t)} \rho(\bm{r}, t)\,d^3x\,.\label{mass}
\end{equation}
The change is straightforwardly evaluated
\begin{equation}
{\frac {dm(t)}{dt}} = \int\limits_{\Omega(t)}\left\{\frac{\partial\rho}{\partial t}
  + \bm{div} (\rho {\bm v})\right\} d^3x\,.\label{dmass}
\end{equation}
Mass is a conserved quantity, ${\frac {dm(t)}{dt}} = 0$, that is, we have a continuity equation for $\rho$:
\begin{equation}\label{continuity} 
\frac{\partial\rho}{\partial t}+ \bm{div} (\rho {\bm v})=0\,.
\end{equation}
Lavoisier (1789) checked this conservation law of mass successfully in experiments.

At the beginning of the 20th century, Newtonian mechanics was supplanted by the special relativity theory (SR) with its four-dimensional (4d) Minkowski space. Accordingly, also a gravitational gauge theory has to take the special-relativistic framework as a starting point. This procedure is reminiscent of Einstein's heuristic derivation of general relativity theory (GR). He started in flat Minkowski space, went over to accelerated frames, and applied subsequently the equivalence principle. Hereby he had to relax the rigidity of the Minkowski space ending up with the Riemannian space of GR. Einstein \cite{Einstein:1921} worked out this procedure in considerable detail in his Princeton lectures of 1921.

Let us stress a point that is often misunderstood. In heuristically deriving a gauge theory of gravity, the physical system under consideration is embedded in a Minkowski space with its rigid 4-parameter translation group $T(4)$ and its rigid 6-parameter Lorentz group $SO(1,3)$. The relaxation of the rigidity of the Minkowski space is achieved by the postulate of {\it local} (instead of rigid) translational and, at a later stage, of Lorentz invariance. The appropriate geometrical framework of spacetime is induced by the gauge principle alone.

We should recall that the Minkowski geometry of SR and the corresponding group of motion, the semidirect product of the translation and the Lorentz groups, the Poincar\'e group $P(1,3) = T(4) \rtimes SO(1,3)$, is supported by all high-energy experiments with great accuracy. Moreover, Wigner (1939) has shown that all elementary quantum mechanical objects obey a {\it mass-spin classification} \cite{Wigner:1939cj}; massless particles are classified according to mass-helicity. The particle attributes mass $m$ and spin $s$ correspond in field theory to the energy-momentum current $\mathfrak T$ and the spin (angular-momentum) current $\mathfrak S$. Accordingly, the mass density of Newton's theory translates field-theoretically into the energy-momentum current of matter: $m \rightarrow {\mathfrak T}$. Thus, the energy-momentum current of matter must be the source of gravity. Later, in Sec.~4 we will see that additionally also the spin current may play a role: $s\rightarrow {\mathfrak S}$. Jointly with the substitution $m \rightarrow {\mathfrak T}$, the mass conservation theorem \eqref{continuity} is dissolved and the conservation of the energy-momentum current replaces it:
\begin{equation}\label{energy}
\partial_j {\mathfrak T}_i{}^j=0\,.
\end{equation}
The nuclear explosion of Alamogordo (1945) is an unmistakable proof of
the violation of the mass conservation law.

By Noether's theorem, energy-momentum conservation is induced by translational invariance of the Lagrangian of an isolated system. Hence without further ado, we can now rephrase the title of the Yang-Mills paper for gravity as follows: {\it Conservation of the   energy-momentum current and translational gauge invariance.} In this way we recognize that the translation group $T(4)$ is the gauge group of ordinary gravity. This point was already made in the beginning of 1960s by many well-known physicists:
\begin{itemize}
\item Sakurai (1960): ``...there exists a deep connection between
  energy conservation and the very existence of the gravitational
  coupling. The gravitational field, being the dynamical manifestation
  of energy, is to be coupled to energy-momentum density....hence the
  gravitational field can interact with itself in the same way as the
  $T=1$ Yang-Mills $B_\mu^{(T)}$ field (which is the dynamical manifestation
  of isospin) can interact with itself.'' \cite{Sakurai:1960} ($T$ denotes the isospin.)

\item Glashow \& Gell-Mann (1961): ``...if we set up the Einstein
  theory by gauge methods then the conclusions are slightly
  different. Instead of an isotopic rotation, we perform a
  4-dimensional translation at each point of space...'' \cite{Glashow:1961} 

\item Feynman (1962): ``The equations of physics are invariant when we
  make coordinate displacements [by] any constant amount $a^\mu$...it
  is possible to investigate how we might make the equations of
  physics invariant when we allow space dependent {\it variable}
  displacements...'' \cite{Feynman}
\end{itemize}
Isn't this clear enough? We propose to consider those three statements
as support for our point of view.\bigskip

\subsubsection*{The energy-momentum current in exterior and in tensor calculus}

The canonical energy-momentum tensor density ${\mathfrak T}_\a{}^\b$ of tensor calculus and the canonical energy-momentum current 3-form of exterior calculus $\Sigma_\a$ are equivalent. We have
\begin{equation}\label{canEnergy}
  \Sigma_\a = {\mathfrak T}_\a{}^\b \epsilon_\b\,,\qquad
  {\mathfrak T}_\a{}^\b =\,^\diamond(\vartheta^\b\wedge\Sigma_\a)\,,
\end{equation}
with
$\epsilon_\b = {\frac 1{3!}}\epsilon_{\b\mu\nu\rho}\vt^\mu\wedge\vt^\nu\wedge\vt^\rho$ and $\epsilon_{\b\mu\nu\rho}$ as the totally antisymmetric Levi-Civita symbol with values $(0,\pm 1)$. The `diamond dual' $^\diamond$ is defined by means of the metric-free Levi-Civita symbol \cite{Birkbook}, and $\vartheta^\beta$ denotes the coframe, see Appendix 1. The different components of ${\mathfrak T}_\a{}^\b$ carry the following physical interpretations:
\begin{equation}\label{componentsEnergy}
  {\mathfrak T}_\a{}^\b=\begin{pmatrix}{\mathfrak T}_0{}^0\!\sim
    \text{energy density}&
    {\mathfrak T}_0{}^b\!\sim\text{energy flux density}\\
    {\mathfrak T}_a{}^0\!\sim \text{momentum density\ } & {\mathfrak T}_a{}^b\!\sim
    \text{momentum flux density\,=\,stress}\end{pmatrix},
\end{equation}
see, e.g., Rezzolla and Zanotti \cite{Rezzolla:2013} or \cite{Birkbook}.

In the hydrodynamic approximation of a relativistic continuum, the non-interacting (dust) matter elements carry momentum density $p_\a$, and the energy-momentum tensor reads ${\mathfrak T}_\a{}^\b = p_\a u^\b$, with the 4-velocity vector field $u^\b$. We immediately recognize a direct analogy between the energy-momentum ${\mathfrak T}_\a{}^\b = p_\a u^\b$ and the Yang-Mills current $J^i{}_I = u^iq_I$:
\begin{equation}\label{chargemass}
({\rm electric\ charge}) \longleftrightarrow ({\rm color\ charge})
\longleftrightarrow ({\rm momentum}).
\end{equation}
Later Sciama \cite{Sciama} developed ``The analogy between charge and spin in general relativity'' (the title of his paper) even further within a gauge approach to gravity. There the natural material source of the gravitational field is a continuum of non-interacting elements with momentum $p_\a$ and internal angular momentum (spin) $s_{\a\b} = -\,s_{\b\a}$.  The corresponding Poincar\'e matter current encompasses the energy-momentum and the spin tensor densities
\begin{equation}\label{Pcurrent}
  \left\{ {\mathfrak T}_\a{}^\b = p_\a u^\b,\>\; {\mathfrak S}_{\a\b}{}^\c
    = s_{\a\b}\, u^\c\right\}.
\end{equation}
In exterior language, the motion of a relativistic continuum is described by the flow 3-form $u = u^\a\epsilon_\a$, and the Poincar\'e currents for the spinning dust matter read
\begin{equation}
  \left\{\Sigma_\a = p_\a\, u,\quad \tau_{\a\b}
    = s_{\a\b}\, u\,\right\}.\label{TSdust}
\end{equation}
 
A priori, $\Sigma_\a$ and ${\mathfrak T}_\a{}^\b$ have 16 independent components. If a {\it metric} $g_{\a\b}$ is available---and this is always the case for a Minkowski space we started with---we can lower the second index and define a tensor of type (0,2):
\begin{equation}\label{lower}
  {\mathfrak T}_{\a\b}:=g_{\b\g}\,{\mathfrak  T}_\a{}^\g\,.
\end{equation}
It can be decomposed into symmetric and antisymmetric pieces according to $16=10\,\oplus \,6$: ${\mathfrak T}_{\a\b}= {\mathfrak T}_{(\a\b)}+ {\mathfrak T}_{[\a\b]}$. Furthermore, the trace can be extracted from the symmetric piece $\check{{\mathfrak T}}_{\a\b}:={\mathfrak T}_{(\a\b)}-\frac 14 g_{\a\b}{\mathfrak T}_\g{}^\g$, with $\check{{\mathfrak T}}_{[\a\b]}=0$ and $\check{{\mathfrak T}}_\g{}^\g=0$. Thus, we arrive at the following decomposition of the canonical energy-momentum tensor:
\begin{equation}\label{Tdecomp}
   {\mathfrak T}_{\a\b}=\check{{\mathfrak T}}_{\a\b} + {\mathfrak T}_{[\a\b]}
   +\frac 14 g_{\a\b}{\mathfrak T}_\g{}^\g\,,\qquad 16=9\oplus 6\oplus1\,.
\end{equation}

In his deduction of GR, Einstein considered as a model for `matter' the classical Euler fluid and the electromagnetic field in vacuum; in  the former case ${\mathfrak T}_{(\a\b)}$ is sufficient, in the latter one $\check{{\mathfrak T}}_{\a\b}$. Accordingly, in GR matter is described by the symmetric energy-momentum tensor ${\mathfrak T}_{(\a\b)}$. If matter with spin is involved, the canonical tensor ${\mathfrak T}_\a{}^\b$ is indispensable. We will come back to the material currents in Sec.~4.3.

\section{Translational gauge theory (TG)}

As we saw in the last section, at the beginning of the 1960s it was already clear to Sakurai, Glashow \& Gell-Mann, and to Feynman that a gravitational gauge theory should be based on translation (or displacement) invariance. Accordingly, the task was to investigate the conservation of the material energy-momentum current and the related invariance under rigid and, subsequently, under local translations. The localization of the translational invariance should create the gravitational field!

Soon thereafter, in the 1970s, a translational gauge theory (TG) was set up. It turned out to be a teleparallelism theory \cite{Itin:2001,Hehl:2016glb}. We delineated the historical development and an up-to-date formalism already in Blagojevi\'c \& Hehl \cite{Blagojevic:2013xpa}.\footnote{See in particular the pages 195, 236, and 241 to 249.}

The paper of Cho \cite{Cho:1975dh}, see also \cite{Nitsch:1979qn}, may be taken as a concise description of a translational gauge theory of gravity. Its structure is revisited from a modern geometrical point of view in the more recent papers of Obukhov and Pereira \cite{Obukhov:2002tm,Pereira:2019woq}, see also \cite{Koivisto:2019ejt}. We abstain from publishing once more this well-known formalism of TG, but refer to the literature \cite{Aldrovandi:2013} instead.

Let us recall that rigid translational invariance is made {\it local} at the price of introducing 4 translational gauge potentials---the coframe $\vartheta^\a = e_i{}^\a dx^i$---which compensate the violation of the rigid invariance: 
\begin{equation}
  \text{rigid transl.\ inv.}\ \stackrel{\text{heur.\
    princ.}}{\longrightarrow}\ \text{local transl.\ inv.\ }{\longrightarrow}
 \text{  coframe $\vartheta^\alpha$ compensates.}
\end{equation}
Thus $\vartheta^\alpha$ is the analog of $A_i{}^I$ above. The curl of $\vartheta^\alpha$, the torsion, $T^\alpha$ ---the analog of $F_{ij}{}^I$ above---, arises as the gravitational field strength,
\begin{equation}\label{torsion}
T^\alpha := D\,\vartheta^\alpha= d\vartheta^\alpha
+\Gamma_{\b}{}^\alpha\wedge \vartheta^{\b}\,,
\end{equation}
with $\Gamma_\a{}^\b$ as the Lorentz connection. The corresponding
curvature vanishes:
\begin{equation}\label{vanishingCurv}
  R_\a{}^\b := d\Gamma_\a{}^\b -\Gamma_\a{}^\g\wedge \Gamma_\g{}^\b = 0\,.
\end{equation}
This signifies that a vector, for instance, can be parallelly transported around in an integrable way. We have a distant parallelism, a teleparallelism. It takes place in a so-called Weitzenb\"ock geometry \cite{Weitzenboeck:1923,Weitzenboeck:1928}, see also \cite{Einstein:1928,Moller:1961jj,Moeller:1961MatFys}.

Analogously to the Yang-Mills case, $T^\alpha \ne 0$ is the criterion for the emerging of a new non-trivial gravitational/translational gauge field. It can be shown that the teleparallelism theory, for a suitable Lagrangian quadratic in the torsion, is equivalent to general relativity of 1916, provided a {\it symmetric} energy-momentum tensor is chosen, see \cite{Erice:1979}. This is, in our opinion, a major achievement which demonstrates that translational gauging leads, via a Weitzenb\"ock spacetime, directly to general relativity with its Riemannian spacetime. In the subsequent Sec.~\ref{PG}, it will turn out that TG is a special case of a Poincar\'e gauge theory of gravity (PG) that we will discuss in quite some detail.

\begin{figure}[!htb]
\centering
\includegraphics[width=1.00\textwidth]{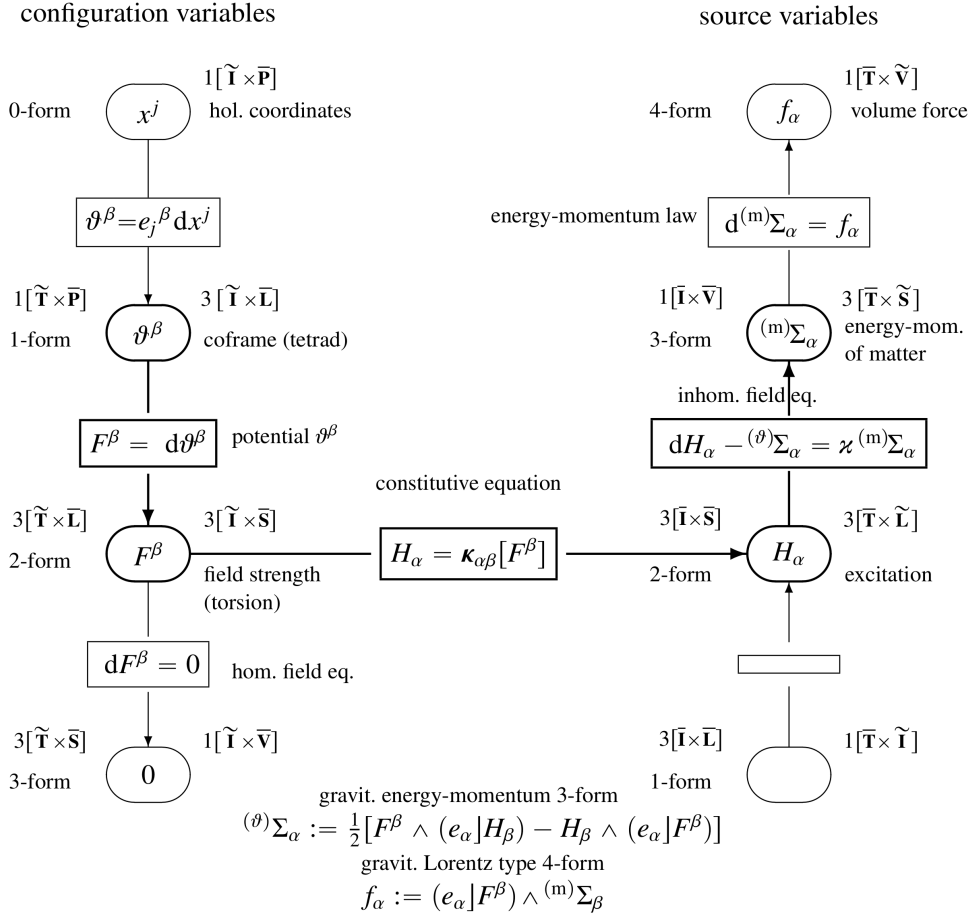}
\caption{Patterned after Tonti \cite{Tonti}, pages 402 and 315. We denoted here the torsion 2-form with $F^\beta$ in order to underline its function as a translational gauge field strength; for full notation
details see \cite{Hehl:2016glb} and \cite{Tonti}. \hfill $^\star$Also known as translation gauge theory of gravity}
\label{TontiTG}
\end{figure}

In order to get a bird's eye of view on TG, we would like to display the structure of the Lagrange-Noether formalism of TG in a Tonti-diagram \cite{Itin:2018dru}, see Fig.~\ref{TontiTG}. The left column is built up from the {\it configuration variables:} The coordinates $x^i$ (four 0-forms), the coframe $\vt^\a$ (four 1-forms), the torsion $T^\b$ (four 2-forms, in the figure called $F^\b$), and eventually $DT^\b\stareq dT^\b$ (four 3-forms). In the right column which depicts {\it source variables,} we start with the 4-forms, the volume force $f_\a$ (four 4-forms), continue with the material energy-momentum $^{{\rm m}}\Sigma_\a$ (four 3-forms) and end up with the translational excitation $H_\a$ (four 2-forms). The corresponding four 1-forms should be the potential of the excitation, but we don't know much about such a quantity; compare, however, with Rund \cite{Rund:1979ea}. The {\it constitutive relation} links the configuration variable 2-form in a linear way to the source variable 2-form, much like in electrodynamics $F=({\bm E}, {\bm B})$ is related linearly to ${H=(\cal D,H})$.

It is remarkable that the {\it configuration} variables are exclusively {\it premetric} concepts, that is, whereas the existence of a linear connection $\Gamma_\a{}^\b$ is necessary, a metric does not enter anywhere.\footnote{Enzo Tonti disagrees on this point since in his way of setting up the basic definitions of his  configuration and source variables, the existence of a Euclidean metric is assumed a priori. We, however, take the view that a suitable premetric generalization is possible.} The analogous is true for the {\it source} variables. In the constitutive laws, however, in which the field strength 2-forms as configuration variables are related to the excitation 2-forms as source variables, a metric tensor is indispensable. This is a lesson which one can take over from the premetric version of electrodynamics, see Post \cite{Post:1962}.

Note that the motion of a point particle in TG is described by the teleparallel analog of the Lorentz force in the Yang-Mills theory (\ref{LFYM})
\begin{equation}
f_i = p_\a F_{ij}{}^\a u^j.\label{LFTG}
\end{equation}
One can prove that the corresponding equation of motion turns out to be the usual geodesic curve \cite{Itin:2016nxk}.

We know that the Minkowski space of SR is an {\it affine space,} see Kopczy\'nski \& Trautman \cite{Kopczynski:1992}, that is, ``a vector space which has lost its origin.'' A translation is an affine concept unrelated to a metric. Consequently, the gauging of translations happens in an affine space with the canonical energy-momentum 3-form $^{({\rm m})}\Sigma_\a$ (16 independent components) as source. No metric is involved at all in this. However, this teleparallelism scheme cannot be directly compared with nature.

In gravity, as we have discussed above, we start with a Minkowski space and apply the gauge procedure with this background. Minkowski space is indispensable as a starting point for treating gravity, as Einstein \cite{Einstein:1921} has taught us.  For defining a symmetric energy-momentum tensor we need a metric, as we saw already in \eqref{lower}. Hence the premetric teleparallelism scheme does not qualify as a bona fide physical theory. However, since we started from SR, we have a metric available and we can use it for formulating the constitutive law of a teleparallelism theory. Then TG becomes the {\it teleparallel equivalent of general relativity} GR$_{||}$, as we discussed above.

\subsubsection*{Why did Einstein arrive in 1915/16 at a Riemann and
  not at a Weitzenb\"ock space? {\it An afterthought}}

\begin{itemize}
\item Einstein (1916) gauged the {\it direction} of a vector.
\item Weyl (1918) gauged the {\it magnitude} (modulus) of a vector;
  however, instead of coupling it to the dilation current (as we know
  today), he coupled it to the unrelated electric current \cite{HMM:1985}.
\item E.Cartan (1923) recognized that Einstein took the flat Minkowski
  space as a vector space instead of an affine space. A Minkowski
  space has no preferred point. But Einstein took in his construction
  a preferred point in order to gauge the direction. In order to get
  rid of this preferred point, Cartan rolled without gliding a
  Minkowski space along a contour of the contorted and curved space
  under consideration (Cartan circuit, see Kr\"oner \cite{Kroner},
  Sharpe \cite{Sharp}, and Sternberg \cite{Sternberg}). This is the
  meaning of this procedure which provides more insight, in our
  opinion, than all those theories using fiber bundles. Fiber bundles
  were successfully applied for internal symmetries, like
  $U(1),\,SU(2), \text{and}\,SU(3)$, but for external, i.e.\ spacetime
  symmetries, they did not provide any further insight. Unfortunately,
  no bundle theorist has essentially contributed to the understanding
  of torsion and/or constructively developed teleparallelism (with the
  possible exception of Sch\"ucking \& Surowitz \cite{Schucking}), as far as we
  can see.
\end{itemize}

\section{Poincar\'e gauge gravity (PG)}\label{PG}

In the next step we will discuss now the gauging of the Poincar\'e group $P(1,3)=T(4) \rtimes SO(1,3)$. Before we do so, we would like to look at a prototypical experiment by Colella, Overhauser, and Werner (COW) on the `behavior' of a neutron beam in a gravitational field. In our understanding of gravity, Newton's apple should nowadays be substituted by a neutron beam as it is used in the COW experiment as quantum system with mass $m$ and spin $s = {\frac \hbar 2}$. As J.~L.~Synge formulated it so beautifully: ``Newton successfully wrote apple = moon, but you cannot write apple = neutron.''

\subsection{Colella-Overhauser-Werner (COW) experiment heralds a new era in gravitational physics: the Kibble laboratory}

\begin{figure}[!htb]
\centering
\includegraphics[scale=0.6]{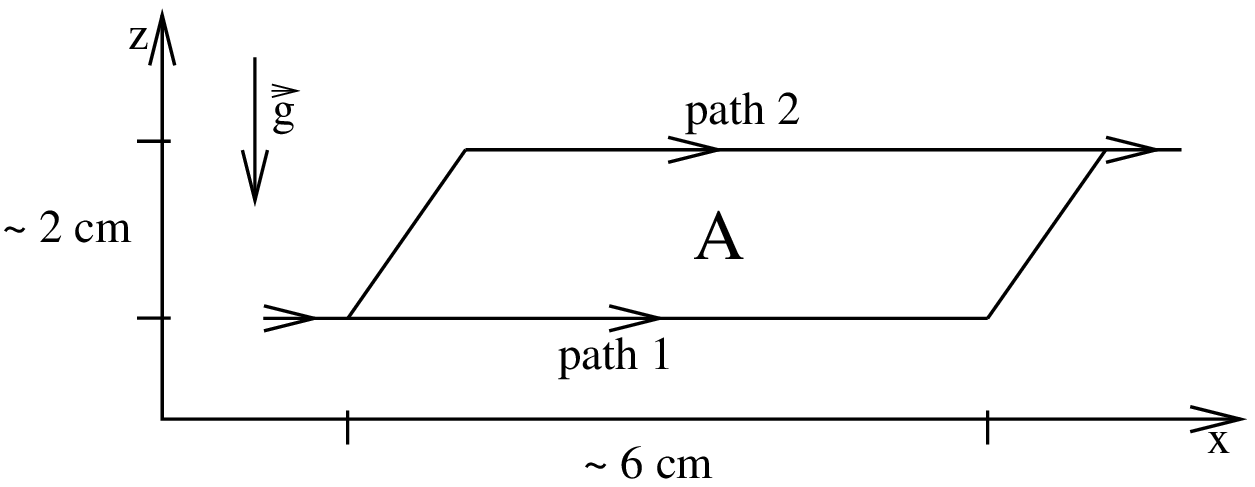}
\caption{COW experiment schematically: A neutron beam is split into two beams
which travel in different gravitational potentials. Eventually the two beams are
reunited and an interferometric picture is observed due to their relative phase shift.}\label{COW}
\end{figure}

\begin{table}
\begin{footnotesize}
  \caption{{\bf Einstein's approach to GR versus the gauge approach to gravity.} Used are a mass point $m$ or a Dirac matter field $\Psi$ (referred to a local frame), respectively. IF means inertial frame, NIF non-inertial frame. The table refers to special relativity up to the second horizontal double line. Below, gravity will be switched on.  Note that for the Dirac spinor already the force-free motion in an {\it inertial} frame {\it does depend} on the mass parameter $m$. The tilde $\tilde{}$ denotes the quantities in a Riemannian spacetime.}\label{SR}
\hspace{-1cm}
\begin{tabular}{|c|c|c|}
\hline
& & \\  
 & {\bf Einstein's approach:} & {\bf Gauge approach:} \\
&Einstein laboratory&  Kibble laboratory\\
  & & \\
\hline
 & & \\
Elementary object in & mass point $m$ & Dirac spinor $\Psi(x)$ of mass $m$ \\
Special Relativity (SR) & with velocity $u^i$ & (with four components) \\
\hline
 & & \\  
Inertial  &  Cartesian coord.\ system $x^i$ & holonomic orth.\ frame \\
frame (IF)  &  $ds^2\stareq o_{ij}\,dx^idx^j$ & 
$e_\alpha=\delta_\alpha^i\,\partial_i,\quad e_\alpha\!\cdot e_\beta=o_{\alpha\beta}$ \\
\hline 
 & & \\
Force-free motion in IF &  ${\dot u}{}^i\stareq 0$ & $(i\gamma^i\partial_i-m)\Psi\stareq 0$ \\ 
\hline
 & & \\  
Non-inertial & arbitrary curvilinear & anholonomic orth. frame $e_\alpha\! =e^i{}_\alpha\partial_i$ \\
frame (NIF)  & coord.\ system $x^{i'}$ & or coframe $\vartheta^\alpha=e_i{}^\alpha dx^i$ \\
\hline
 & & \\  
Force-free & ${\dot u}{}^{i}+u^{j}u^{k}\tilde{\Gamma}_{jk}{}^i = 0$ & $\left[i\gamma^\alpha e^i{}_\alpha(\partial_i+\Gamma_i) - m\right]\!\Psi=0 $ \\
motion in NIF &  &   $\Gamma_i:={\frac 12}\,\Gamma_i{}^{\beta\gamma}\rho_{\beta\gamma}$\qquad Lorentz\\
\hline
 & & \\  
Non-inertial & $\tilde{\Gamma}_{jk}{}^i$ & $\vartheta^\alpha,\quad \Gamma^{\alpha\beta} =-\Gamma^{\beta\alpha}$ \\
geometrical objects & 40  &  16\qquad +\qquad  24 \\
\hline 
 & & \\  
Constraints & ${\tilde R}^{\alpha\beta}(\partial\tilde{\Gamma},\tilde{\Gamma})=0$ & 
$T^{\alpha}(\partial e,e,\Gamma)\!=\!0,\ R^{\alpha\beta}(\partial\Gamma,\Gamma)\!=\!0$ \\
in SR   & 20 &     24\qquad + \qquad 36 \\
\hline
 & & \\  
Global IF & $g_{ij}\stareq o_{ij}\,,\quad \tilde{\Gamma}_{jk}{}^i\stareq 0$
 & $\left(e_i{}^\alpha,\ \Gamma_i{}^{\alpha\beta}\right)\stareq\left(\delta^\alpha_i,0\right)$ \\
\hline
\hline
 & & \\  
Archetypal experiment & Apple in grav. field (Newton) & Neutron in grav. field (COW) \\
 & & \\  
\hline
 & & \\  
Switch on & ${\tilde R}^{\alpha\beta}\neq 0$ & $T^{\alpha}\neq 0,\quad R^{\alpha\beta}\neq 0$ \\
gravity & Riemann spacetime & Riemann-Cartan spacetime\\
\hline
 & & \\  
Local IF (`Einstein elevator') & $g_{ij}\vert_P \stareq o_{ij}\,,\quad \tilde{\Gamma}_{jk}{}^i\vert_P
\stareq  0$ & $(e_i{}^\alpha,\ \Gamma_i{}^{\alpha\beta})\vert_P \stareq (\delta^\alpha_i,0)$ \\
\hline
 & & \\
Gravitational & ${\rm {\tilde Ric}\!-\!{\frac 12}tr({\tilde Ric})}\sim$ mass & ${\rm Ric -
{\frac 12}tr(Ric)}\sim$ mass \\ 
field &  & ${\rm Tor+2\ tr(Tor)}\sim$ spin \\
equations & {\bf GR} & {\bf EC} \\
\hline
\end{tabular}
\end{footnotesize}
\end{table}

The quantum mechanical properties of a neutron wave/particle in interaction with the ordinary Newtonian gravitational field were first observed in the Colella-Overhauser-Werner neutron interferometer\footnote{The interferometer, built from a silicon monocrystal, had a linear size of about 10 cm. In the energy range covered by the COW experiment, the neutron can be considered to be elementary, that is, its quark structure can be neglected.} in 1975 \cite{Colella:1975dq}, see also Rauch \& Werner \cite{RauchWerner} for a corresponding textbook presentation. The scheme of the experiment is sketched in Fig.~\ref{COW}. Later more accurate measurements with atomic beams were performed by Kasevich \& Chu \cite{Kasevich:1991zz}, see also the most recent developments by Asenbaum et al.\ \cite{Asenbaum:2017rwf} and Overstreet et al.\ \cite{Overstreet:2017gdp}. Unfortunately, in most textbooks on GR, these important experiments are not even mentioned and the old Einstein procedure of using Newton's apple for a heuristic derivation of GR is simply copied.

To predict the result of the COW experiment was not complicated: provided the neutron spin is not polarized, one couples the Schr\"odinger equation for a neutron to the Newtonian gravitational potential. The quantum phase shift predicted, and experimentally confirmed by COW and Kasevich \& Chu, was mass dependent. Then it was soon argued that this would violate the equivalence principle. However, this interpretation turned out to be incorrect, see Audretsch et al.\ \cite{Audretsch:1992tg}. It was necessary to take the wave function of the neutron---in generalization of the Newtonian point particle---as a new basic ingredient for the discussion of the equivalence principle: the neutron wave/particle of the COW-experiment supplanted the Newtonian point particle moving in a gravitational field.\footnote{For a very down-to-earth and highly interesting discussion of the interaction between a quantum system and a classical gravitational field, one should compare Nesvizhevsky \& Voronin \cite{NesvizhevskyVoronin}.} Here quantum mechanical results for the  test matter are needed for the foundation of the theory of gravity.

Einstein \cite{Einstein:1921} heuristically derived GR from SR by considering mass points and electromagnetic fields in an accelerating and thus in a noninertial frame of reference. In the {\it Einstein laboratory}, in which he executed his thought experiments, the acceleration was described by using curvilinear coordinates $x^i$, with $i=0,1,2,3$. The inertial forces, which emerge in an accelerated reference system, were locally equivalent to the corresponding gravitational forces; for a detailed discussion see, for instance, Audretsch et al.\ \cite{Audretsch:1992tg} or Blagojevi\'c and Hehl \cite{Blagojevic:2013xpa}, in particular Fig.\ 4.1 therein, and our Table~\ref{SR}.

In short, we can characterize the Einstein lab(oratory) as follows: Specified are \vspace{-5pt}
\begin{description}
\item[{\rm \{E1\}}]a neutral point particle with mass $m$;
\item[{\rm \{E2\}}] an inertial frame $K$ span by Cartesian coordinates;
\item[{\rm \{E3\}}] an accelerated (i.e., non-inertial) frame $K'$ span by
  curvilinear coordinates;
\item[{\rm \{E4\}}] a homogeneous gravitational field described by $g_{ij}$
  referring to the curvilinear coordinates; and
\item[{\rm \{E5\}}] light rays.
\end{description}

Einstein's procedure, by means of which he deduced GR, could have a broader domain of application than that encompassed by the tools and constructs employed in setting up the theory. Still, it is hard for us to believe that Einstein's discussion would also cover the COW experiment. After all, the neutron has quantum mechanical properties, it is a fermion, and it has spin $s = {\frac \hbar 2}$.

Accordingly, we consider the {\it Kibble laboratory}, in which a fermion is described by a Dirac wave function with respect to a local reference (co)frame (vierbein) $e_i{}^\a$, where $\a=0,1,2,3$ numbers the frame vectors, see Kibble \cite{Kibble:1961ba}.  As soon as the reference frames are accelerated, they are no longer aligned and pick up a non-vanishing curl of these frames, the so-called object of anholonomity. In a non-relativistic approximation and neglecting its spin, the neutron obeys the stationary Schr\"odinger equation in the external homogeneous Newtonian gravitational field. If one solves this equation, the experimentally observed gravitational phase shift is described successfully.

The rationale of all of this is that we should simply study, in Minkowski space, a Dirac field in a Kibble lab, that is, in an accelerated frame of reference $e_i{}^\a(x)$ with $\partial_{[i}e_{j]}{}^{\a}\ne 0 $, and read off its inertial forces. This is what we did, see \cite{Hehl:1990nf} and the literature quoted there. Subsequently, we executed a Foldy-Wouthuysen transformation in order to determine the non-relativistic limit of the Dirac equation. If one neglects the spin, one recovers the COW term thereby certifying the correctness of this procedure. If the spin is kept, as one should do since the fundamental building blocks of our universe are fermions, one recovers additionally a spin-rotation coupling predicted earlier by Mashhoon \cite{Mashhoon:1988zz,Mashhoon:1997qc} and experimentally confirmed in the meantime, see Danner et al. \cite{Danner:2020}.

Thus, the Kibble lab(oratory) can be described as follows: it contains
\begin{description}
\item[{\rm \{K1\}}] an unquantized Dirac spinor\footnote{It is demonstrated in the textbooks of Ivanenko \& Sokolov \cite{Iwanenko:1953} and Hund \cite{Hund:1954} that it is useful at relatively low energies to consider, in an approximate way, the Dirac spinor as a {\it  classical} field---in the so-called first quantized version. Thus, quantum field theory is excluded from our considerations. In this context, one should also compare the discussion of Cao \cite{Cao:2019} in his Sec.11.3: One is searching for a background independent quantum theory of gravity, but a Riemann-Cartan background \`a la Kibble et al.\ seems to be a reasonable starting point.} (fermionic field with mass $m$  and spin $s= {\frac \hbar 2}$);
\item[{\rm \{K2\}}] an inertial frame, $\vt^\a=\delta^\a_i dx^i$ spanned by
  Cartesian coordinates;
\item[{\rm \{K3\}}] a translationally and rotationally accelerated frame,
  $\vt^{\a '}$ spanned by an arbitrary orthonormal frame;
\item[{\rm \{K4\}}] homogeneous gravitational fields described by $(e_i{}^\a,\Gamma_i{}^{\a\b})$; and
\item[{\rm \{K5\}}] light rays.
\end{description}
If we compare the different labs, we recognize that (i) the objects considered are different, (ii) the notions of inertial systems are different, and (iii), see Mashhoon \cite{Mashhoon:2003ax}, the rotational acceleration plays an additional role in the Kibble lab. The rest, in particular the application of the equivalence principle, is similar.  Because we consider spinors in the Kibble lab, we get, by using Einstein's original ideas, a modified outcome.

Conventionally, the equivalence principle is only discussed in the Einstein laboratory. The Kibble laboratory, which is really based on \'Elie Cartan's moving (co)frames (rep\`eres mobiles), is, in our opinion, a necessity if Dirac particles are considered. This amounts to a generalization of the equivalence principle to a `more local' neighborhood, see, in particular von der Heyde \cite{vdH:1975}---the curvilinear coordinates $x^i$ are generalized to arbitrary orthonormal coframes $\vt^\a = e_i{}^\a dx^i$.

According to the equivalence principle, the Riemann-Cartan (RC) spacetime looks Minkowskian from a local point of view. In a RC-spacetime, at any fixed point with coordinates ${\stackrel{\circ}{x}}{\,}^k$ it is possible to trivialize the gravitational gauge potentials \cite{vdH:1975}:
\begin{equation}\label{238}
\left.
\begin{aligned}
  e_i{}^{\alpha}\,\vline\,{\hbox{\raisebox{-2ex}{\scriptsize $x^k\!={\stackrel{\circ}{x}}{\,}^k$}}}
  &\stackrel{\ast}{=}\delta^\alpha_i\  \\
  \Gamma_i{}^{\alpha\beta}\,\vline\,{\hbox{\raisebox{-2ex}{\scriptsize $x^k\!={\stackrel{\circ}{x}}{\,}^k$}}}
  &\stackrel{\ast}{=} 0\
\end{aligned}
\right\}.
\end{equation}
This is important to recognize: In spite of the presence of torsion in a RC-spacetime, at any fixed point, the local connection $\Gamma_i{}^{\a\b}$ can be transformed to zero \cite{Hartley,Iliev,Nester:2010}. What in an Einstein lab is the geodesic coordinate system becomes the trivialized reference frame (\ref{238}) in a Kibble lab; this is a widely underestimated point. The frame (\ref{238}) in RC-space supersedes the geodesic coordinate system in a Riemannian space.

\subsection{Poincar\'e gauge gravity kinematics}

The standard model of particle physics is based on gauge theories for the internal symmetries $U(1), SU(2), SU(3)$, see O'Raifeartaigh \cite{ORaifeartaigh:1978jea}. Accordingly, apart from GR, the gauge idea seems to underlie all physical theories. However, already fairly early also gravity was understood as a gauge theory. It was Utiyama \cite{Utiyama:1956} who paved the way in this direction by using the Lorentz group $SO(1,3)$ as a gauge group for gravity. It turned out to be unsuccessful, though, since the current coupling to the Lorentz group is the angular momentum current. However, as we know from Newton's theory of gravity, it is the mass density or---according to SR---the energy-momentum current that gravity has as its source. And energy-momentum couples to the translation group $T(4)$.

Flat gravity-free Minkowski space has the Poincar\'e group $T(4)\rtimes SO(1,3)$, the semi-direct product of the four-parameter translation group\footnote{A highly original contribution to the understanding of translation gauge invariance was provided by Tresguerres \cite{Tresguerres:2007ih}.} $T(4)$ and the six-parameter Lorentz group $SO(1,3)$, as its group of motions. Accordingly, Minkowski space is invariant under rigid (`global') Poincar\'e transformations.  Consequently, as found by Wigner \cite{Wigner:1939cj}, a quantum mechanical system in a Minkowski space can be classified according to {\it mass and spin.} The corresponding field-theoretical currents are the material energy-momentum and spin angular momentum currents $\Sigma_\alpha$ and $\tau_{\a\b} = -\,\tau_{\b\a}$, respectively.

Thus, if we want to apply Einstein's recipe for setting up a gravitational theory based on the equivalence principle, we have to introduce accelerated frames in Minkowski space. Due to the involvement of a quantum mechanical system, see the COW neutrons or the Kasevich \& Chu atoms mentioned above, we have to turn to Kibble's laboratory and to introduce coframes as reference systems. This yields, as was shown by Sciama \cite{Sciama} and Kibble \cite{Kibble:1961ba}, via {\it local} Poincar\'e invariance a Riemann-Cartan spacetime with torsion $T_{\a\b}{}^\g$ and curvature $R_{\a\b}{}^{\g\d}$. The rigid Lie algebra of the Poincar\'e group is extended to a so-called deformed, soft, or local ``Lie algebra'' ($D_\a$ and $\rho_{\a\b} = -\,\rho_{\b\a}$ generate translations and Lorentz transformations, respectively):
\begin{equation}\label{soft}
\left.
\begin{aligned}
\null  [D_\a,D_\b] &= -\,T_{\alpha\beta}{}^{\gamma}
  D_{\gamma}+R_{\alpha\beta}{}^{\gamma\delta}
  \rho_{\delta\gamma}\  \\
  [\rho_{\alpha\beta},D_\gamma] &= -\,g_{\gamma\alpha}D_{\beta} + g_{\gamma\beta}D_{\alpha}\ \\
  [\rho_{\alpha\beta},\rho_{\mu\nu}] &= -\,g_{\alpha\mu}\rho_{\beta\nu} + g_{\alpha\nu}\rho_{\beta\mu}
  + g_{\beta\mu}\rho_{\alpha\nu} - g_{\beta\nu}\rho_{\alpha\mu}\
\end{aligned}
\right\}.
\end{equation}
The rigid Lie algebra of Minkowski space is recovered for $T_{\a\b}{}^\g=0$ and $R_{\a\b}{}^{\g\d}=0$; then, in Cartesian coordinates, $D_\a\rightarrow \partial_a$, for details see \cite{Erice:1979}.\footnote{This local Lie algebra structure has also   been found in the gauge theory of the de Sitter group $SO(2,3)$, see  Stelle \& West \cite{Stelle}. Note that the cosmological model favored by present observations seems to favor an underlying anti-de Sitter universe.}  Thus the Riemannian spacetime of GR is generalized to the Riemann-Cartan spacetime of the Poincar\'e gauge theory. And the underlying reason for this generalization is evident: It is the application of the Einstein procedure to a quantum mechanical system instead of to a classical point particle. The method remains the same, the objects to which it is applied to were generalized. Instead of an Einstein laboratory with curvilinear coordinates, we use a Kibble lab with frames in order to encompass also fermionic fields.

Up-to-date reviews of the Poincar\'e gauge theory of gravity can be found in \cite{Obukhov:2006,Obukhov:2018}, and for more historic and technical details readers may refer to \cite{Blagojevic:2013xpa,Ponomarev,Mielke,Erice:1979,PRs}. Here we briefly outline the most essential notions and constructions. 

Following the general Yang-Mills-Utiyama-Sciama-Kibble gauge-theoretic sche\-me, the 10-parameter Poincar\'e group $T_4\,\rtimes\,SO(1,3)$ gives rise to the 10-plet of the gauge potentials which are consistently identified with the coefficients $e_i{}^\a$ of the orthonormal coframe $\vartheta^\alpha = e_i{}^\a dx^i$ (4 potentials corresponding to the translation subgroup $T_4$) and the components $\Gamma_{i}{}^{\alpha\beta} = -\,\Gamma_{i}{}^{\beta\alpha}$ of the local connection $\Gamma^{\alpha\beta} = \Gamma_{i}{}^{\alpha\beta} dx^i$ (6 potentials for the Lorentz subgroup $SO(1,3)$). The corresponding covariant curls, the field strengths of translations and Lorentz rotations, 
\begin{align}
T_{ij}{}^\alpha &= \partial_i e_j{}^\a - \partial_j e_i{}^\a
+ \Gamma_{i\beta}{}^\alpha e_j{}^\b - \Gamma_{j\beta}{}^\alpha e_i{}^\b,\label{Tor}\\ \label{Cur}
R_{ij}{}^{\alpha\beta} &= \partial_i\Gamma_j{}^{\alpha\beta} - \partial_j\Gamma_i{}^{\alpha\beta} 
+ \Gamma_{i\gamma}{}^\beta\Gamma_j{}^{\alpha\gamma} - \Gamma_{j\gamma}{}^\beta\Gamma_i{}^{\alpha\gamma},
\end{align}
are the torsion $T_{ij}{}^\a$ and the curvature $R_{ij}{}^{\a\b}$, both antisymmetric in $i$ and $j$. This naturally introduces the Riemann-Cartan geometry \cite{Schrodinger,Schouten:1954,Schouten:1989,Hehl:deBroglie} on the spacetime manifold.

Obviously, both gravitational field potentials transform covariantly as covectors
\begin{equation}\label{eGdiff}
e_{i'}{}^\alpha = {\frac {\partial x^j}{\partial x^{i'}}}\,e_j{}^\alpha,\qquad
\Gamma_{i'}{}^{\alpha\beta} = {\frac {\partial x^j}{\partial x^{i'}}}\,\Gamma_j{}^{\alpha\beta},
\end{equation}
under arbitrary local coordinate transformations (diffeomorphisms) 
\begin{equation}
x^i \longrightarrow x^{i'} = x^{i'}(x^j).\label{diff}
\end{equation}
As a result, the gauge gravitational field strengths (\ref{Tor}) and (\ref{Cur}) transform covariantly
\begin{equation}\label{TRdiff}
T_{i'j'}{}^\alpha = {\frac {\partial x^k}{\partial x^{i'}}}{\frac {\partial x^l}{\partial x^{j'}}}
\,T_{kl}{}^\alpha,\qquad R_{i'j'}{}^{\alpha\beta} = {\frac {\partial x^k}{\partial x^{i'}}}
{\frac {\partial x^l}{\partial x^{j'}}}\,R_{kl}{}^{\alpha\beta},
\end{equation}
as second rank skew-symmetric tensors under the change of coordinates (\ref{diff}).

The action of the local Lorentz group is nontrivial. The Lorentz
transformation, by definition, leaves the metric invariant:
\begin{equation}
\Lambda(x)_\alpha{}^\mu\Lambda(x)_\beta{}^\nu g_{\mu\nu} = g_{\alpha\beta},\label{ginv}
\end{equation}
which means that the Lorentz transformed (``rotated'') translational
potential
\begin{equation}
e'_i{}^{\mu}=\Lambda(x)_\alpha{}^\mu e_i{}^\alpha,\label{cofL}
\end{equation}
remains orthonormal under (\ref{ginv}). The corresponding
transformation law for the local Lorentz connection is inhomogeneous:
\begin{equation}
\Gamma'_{i\alpha}{}^\beta = \Lambda(x)_\nu{}^\beta\Gamma_{i\mu}{}^\nu\Lambda^{-1}(x)_\alpha{}^\mu + 
\Lambda(x)_\mu{}^\beta \partial_i\Lambda^{-1}(x)_\alpha{}^\mu.\label{connL}
\end{equation}
Accordingly, for the torsion and the curvature we find
\begin{equation}
T'_{ij}{}^{\mu} = \Lambda(x)_\alpha{}^\mu T_{ij}{}^\alpha,\qquad R'_{ij}{}^{\alpha\beta} = 
\Lambda(x)_\mu{}^\alpha\Lambda(x)_\nu{}^\beta R_{ij}{}^{\mu\nu}.\label{torcurL}
\end{equation}

An infinitesimal Lorentz transformation 
\begin{equation}
\Lambda(x)_\alpha{}^\mu = \delta_\alpha^\mu + \varepsilon(x)_\alpha{}^\mu\label{infL}
\end{equation}
is described by the parameters $\varepsilon(x)_\alpha{}^\mu$ which, in
view of (\ref{ginv}), satisfy
\begin{equation}
\varepsilon_{\alpha\beta} + \varepsilon_{\beta\alpha} = 0.\label{eps}
\end{equation}
Noting that $\Lambda^{-1}(x)_\alpha{}^\mu = \delta_\alpha^\mu - \varepsilon(x)_\alpha{}^\mu$, we derive the infinitesimal form of transformation laws for the coframe (\ref{cofL}) and the connection as (\ref{connL}):
\begin{align}
\delta e_i{}^\a &= \varepsilon_\b{}^\a e_i{}^\b,\label{infT}\\ 
\delta\Gamma_i{}^{\alpha\beta} &= -\,\partial_i\varepsilon^{\alpha\beta}
- \Gamma_{i\gamma}{}^\alpha\varepsilon^{\gamma\beta}
- \Gamma_{i\gamma}{}^\beta\varepsilon^{\alpha\gamma}.\label{infG}
\end{align}
On account of the explicit generators $(\rho^\mu{}_\nu)_{\alpha\beta} = \delta^\mu_\alpha g_{\nu\beta} -
\delta^\mu_\beta g_{\nu\alpha}$ for the vector representation of the Lorentz group, we can recast (\ref{infT}) into
\begin{equation}
\delta e_i{}^\mu = - \,{\frac 12}\varepsilon^{\alpha\beta}(\rho^\mu{}_\nu)_{\alpha\beta}\,
e_i{}^\nu.\label{infC}
\end{equation}
This can directly be extended to the infinitesimal transformation of any field $\psi^A$ which belongs to an arbitrary representation of the Lorentz group with the generators $(\rho^A{}_B)_{\alpha\beta}$:
\begin{equation}\label{infP}
\delta\psi^A = - \,{\frac 12}\varepsilon^{\alpha\beta}(\rho^A{}_B)_{\alpha\beta}\,\psi^B.
\end{equation}
In accordance with the transformation laws (\ref{infP}) and
(\ref{infG}), the covariant derivative of an arbitrary field is defined by
\begin{equation}\label{Dp}
D_i\psi^A = \partial_i\psi^A - {\frac 12}\Gamma_i{}^{\alpha\beta}(\rho^A{}_B)_{\alpha\beta}\,\psi^B.
\end{equation}

We conclude this section by noticing that the gravitational Poincar\'e
gauge field strengths satisfy the two Bianchi identities:
\begin{align}
D_{[i}T_{jk]}{}^\alpha &= R_{[ijk]}{}^\alpha,\label{Bianchi2c}\\
D_{[i}R_{jk]}{}^{\alpha\beta} &= 0.\label{Bianchi1c}
\end{align}
Here, $R_{ijk}{}^\alpha = R_{ij\beta}{}^\alpha e_k{}^\beta$, and the covariant derivative $D_i$ is defined 
by (\ref{Dp}).

Although the tensor language is, so to say, a mother-tongue for relativists, the use of the modern coordinate free formalism of exterior forms proves to be extremely convenient in the gauge gravity theory. Accordingly, {\sl we switch here to exterior forms.} Then the first (\ref{Bianchi2c}) and the second (\ref{Bianchi1c}) Bianchi identities read, respectively, $DT^\a=R_\g{}^\a\wedge\vt^\g$ and $D R^{\a\b}=0$.

\subsection{Matter Lagrangian: Currents and conservation laws}

Before we discuss the dynamics, we need to revisit the cornerstones of a gauge theory, i.e.\ the conservation laws.

For the sake of generality, we will assume a matter field $\psi^A$ to be a $p$-form that transforms according to an arbitrary representation of the Lorentz group, cf. (\ref{infP}). Let us consider a general matter Lagrangian 4-form
\begin{eqnarray} 
L = L(\vartheta^\alpha\,, d\vartheta^\alpha\,,\Gamma^{\alpha\beta}\,, 
d\Gamma^{\alpha\beta}\,,\psi^A, d\psi^A) 
= L(\psi^A, D\psi^A, \vartheta^\alpha, T^\alpha, R^{\alpha\beta})\,.\label{L} 
\end{eqnarray}
Note that we take into account a possibility of the nonminimal coupling between matter and the gravitational field by allowing the dependence of $L$ on the Poincar\'e gauge field strengths.  In the minimal coupling scheme, $L = L(\psi^A, D\psi^A, \vartheta^\alpha)$, so that the matter interacts with the gravity only via the Poincar\'e gauge field potentials $(\vartheta^\alpha\,, \Gamma^{\alpha\beta})$ which contribute to the Lagrangian either directly or via the covariant derivatives $D\psi^A = d\psi^A - {\frac 12}\Gamma^{\alpha\beta}\wedge(\rho^A{}_B)_{\alpha\beta}\psi^B$.

Independent variations of the matter and gravitational arguments $\psi^A, \vartheta^\alpha, \Gamma^{\alpha\beta}$ yield for the matter Lagrangian
\begin{eqnarray}
\delta L &=& -\,\delta\vartheta^{\alpha}\wedge\Sigma_\alpha
- {\frac 12}\delta\Gamma^{\alpha\beta}\wedge\tau_{\alpha\beta} 
+\delta\psi^A\wedge{\frac{\delta L}{\delta\psi^A}}\label{vard1}\\
&& + \,d\left[\delta\vartheta^{\alpha}\wedge{\frac
{\partial L}{\partial T^{\alpha}}}+ \delta\Gamma^{\alpha\beta}
\wedge{\frac{\partial L}{\partial R^{\alpha\beta}}}+
\delta\psi^A\wedge{\frac{\partial L}{\partial D\psi^A}}\right]\, .\nonumber
\end{eqnarray}
Here, for a gauge--invariant Lagrangian $L$, the expression
\begin{equation}
{\frac {\delta L} {\delta\psi^A}}
={\frac {\partial  L}{\partial\psi^A}} - (-1)^{p}D\,
{\frac {\partial L}{\partial (D\psi^A)}} \label{psi0}
\end{equation}
is the covariant {\it variational derivative} of $L$ with respect to the 
matter $p$--form $\psi^A$. The {\it matter currents} in (\ref{vard1}) are introduced by
\begin{align}
\Sigma_{\alpha} :=& -\,{\frac {\delta L}{\delta\vartheta^{\alpha}}} =  
- \,{\frac {\partial L}{\partial\vartheta^{\alpha}}} 
- D\,{\frac {\partial L}{\partial T^{\alpha}}}\, ,\label{sigC0}\\
\tau_{\alpha\beta} :=& -\,2{\frac {\delta L}{\delta\Gamma^{\alpha\beta}}} =   
(\rho^A{}_B)_{\alpha\beta}\psi^B\wedge{\frac {\partial L} {\partial (D\psi^A)}}\nonumber\\
& -\,\vartheta_\alpha\wedge {\frac {\partial L}{\partial T^\beta}} +
\vartheta_\beta\wedge {\frac {\partial L}{\partial T^\alpha}} 
-2D{\frac {\partial L}{\partial R_\alpha{}^\beta}}\,.\label{spin0}
\end{align}
These are the energy-momentum current and the spin angular momentum current of matter, respectively. 
Using the master formula (\ref{vard1}), we can derive the conservation laws for them.

The local {\it translations}, or general coordinate transformations (diffeomorphisms), are generated by the Lie derivatives along arbitrary vector fields $\xi = \xi^\alpha e_\alpha$ on the spacetime manifold. When the Lagrangian $L$ is invariant under the local diffeomorphisms, the master formula (\ref{vard1}) gives rise to two identities\footnote{Hint: one obtains two identities because {\it both}, $\xi^\alpha$ {\it and\/} $d\xi^\alpha$, are {\it point-wise arbitrary}.}. One consequence of the diffeomorphism invariance is the {\it first Noether identity}
\begin{eqnarray}
D\Sigma_\alpha \equiv  (e_\alpha\rfloor T^\beta)\wedge\Sigma_\beta
+ {\frac 12}(e_\alpha\rfloor R^{\beta\gamma})\wedge\tau_{\beta\gamma} +\,W_{\alpha},\label{conmomC}
\end{eqnarray}
where the generalized force is $W_{\alpha} := -\,(e_\alpha\rfloor D\psi^A)\wedge{\frac{\delta L}{\delta\psi^A}} - (-1)^p(e_\alpha\rfloor\psi^A)\wedge D{\frac{\delta L}{\delta\psi^A}}$. As another consequence of the translational invariance one finds the explicit form of the {\it canonical energy--momentum current}:
\begin{align}
\Sigma_\alpha =&  \,(e_\alpha\rfloor D\psi^A)\wedge {\frac{\partial L}{\partial D\psi^A}}
+ (e_\alpha\rfloor\psi^A)\wedge{\frac{\partial L}{\partial\psi^A}} - e_\alpha\rfloor L\nonumber\\
& -\,D{\frac{\partial L}{\partial T^\alpha}} + (e_{\alpha}\rfloor T^\beta)\wedge
{\frac{\partial L}{\partial T^\beta}} + (e_{\alpha}\rfloor R^{\beta\gamma})\wedge 
{\frac{\partial L}{\partial R^{\beta\gamma}}}.\label{momC}
\end{align}

Note that the second lines in (\ref{spin0}) and (\ref{momC}) describe the nonminimal coupling contributions to the spin angular momentum current and the canonical energy-momentum current of matter, respectively.

The identity (\ref{conmomC}) is given in the {\it strong} form, without using the field equations. The generalized force $W_\alpha = 0$ vanishes when the $\psi^A$ satisfy the Euler-Lagrange equations ${\frac{\delta L}{\delta\psi^A}} = 0$. Then the Noether identity (\ref{conmomC}) reduces to the conservation law of energy-momentum in the framework of the Poincar\'e gauge theory:
\begin{eqnarray}\label{Tcons}
D\Sigma_\alpha \cong (e_\alpha\rfloor T^\beta)\wedge\Sigma_\beta 
+ {\frac 12}(e_\alpha\rfloor R^{\beta\gamma})\wedge\tau_{\beta\gamma}.
\end{eqnarray}
The sign $\cong$ denotes `on shell,' that is, the field equations are assumed to be fulfilled. On the right-hand side of the identity (\ref{conmomC}) and of the conservation law (\ref{Tcons}) for the canonical energy-momentum current, we find the typical {\it Lorentz-type force} terms. They have the general structure {\it field strength} $\times$ {\it current}.

We assume that the Lagrangian $L$ is invariant under a local Lorentz transformations $\delta\vartheta^{\alpha} = \varepsilon_{\beta}{}^{\alpha}\,\vartheta^{\beta}$, $\delta\Gamma^{\alpha\beta} = - D\varepsilon^{\alpha\beta}$, $\delta\psi^A = -\,{\frac 12}\varepsilon^{\alpha\beta}\,(\rho^A{}_B)_{\alpha\beta}\,\psi^B$ ---recall (\ref{infT}), (\ref{infG}) and (\ref{infP}). Then, from the master formula (\ref{vard1}), we find the {\it second Noether identity}
\begin{equation}\label{Noe2}
  D\tau_{\alpha\beta} + \vartheta_\alpha\wedge\Sigma_\beta
  - \vartheta_\beta\wedge\Sigma_\alpha \equiv W_{\alpha\beta}.
\end{equation}
The generalized torque is defined here as $W_{\alpha\beta} := -\,(\rho^A{}_B)_{\alpha\beta}\psi^B\wedge
{\frac{\delta L}{\delta\psi^A}}$. The latter vanishes when the matter field equation ${\frac{\delta L}{\delta\psi^A}} = 0$ is satisfied, and then (\ref{Noe2}) reduces to the weak conservation law of the total
angular momentum
\begin{equation}\label{Scons}
D\tau_{\alpha\beta} + \vartheta_\alpha\wedge\Sigma_\beta - \vartheta_\beta\wedge\Sigma_\alpha \cong 0.
\end{equation}

\subsection{Gravitational Lagrangian: Noether identities}

We assume that the gravitational Lagrangian 4-form 
\begin{equation}
V = V(\vartheta^{\alpha}, T^{\alpha}, R_{\alpha}{}^{\beta})\label{lagrV}
\end{equation}
is an arbitrary function of the gravitational field variables. Its variation can be computed with the help of the master formula (\ref{vard1}), and we recast the result into
\begin{equation}
\delta V = \delta\vartheta^{\alpha}\wedge{\cal E}_\alpha
+ \delta\Gamma^{\alpha\beta}\wedge{\cal C}_{\alpha\beta} 
+ \,d\left[\delta\vartheta^{\alpha}\wedge H_\alpha + \delta\Gamma^{\alpha\beta}
\wedge H_{\alpha\beta}\right]\label{deltaV}
\end{equation}
by writing the variational derivatives with respect to the Poincar\'e
gauge gravitational potentials as
\begin{align}
{\mathcal E}_\alpha &:= {\frac{\delta V}{\delta\vartheta^{\alpha}}} = 
DH_{\alpha} - E_{\alpha}, \label{dVt}\\ 
{\mathcal C}_{\alpha\beta} &:= {\frac{\delta V}{\delta\Gamma^{\alpha\beta}}} 
= DH_{\alpha\beta} - E_{\alpha\beta}\,.\label{dVG}
\end{align}
Here we introduced the {\it gauge field momenta} 2-forms
\begin{equation} 
H_{\alpha} := {\frac{\partial V}{\partial T^{\alpha}}}\,,\qquad  
H_{\alpha\beta} := {\frac{\partial V}{\partial R^{\alpha\beta}}}\, ,\label{HH}
\end{equation}  
and defined the $3$--forms of the {\it canonical} energy--momentum and
the canonical spin for the gravitational gauge fields:
\begin{equation} 
E_{\alpha} := -\,{\frac{\partial V}{\partial\vartheta^{\alpha}}},\qquad
E_{\alpha\beta} := -\,{\frac{\partial V}{\partial\Gamma^{\alpha\beta}}} = 
- \vartheta_{[\alpha}\wedge H_{\beta]}\,. \label{EE}
\end{equation}

Diffeomorphism invariance of $V$ yields the {\it Noether identities}
\begin{align}
E_{\alpha} &\equiv -\,e_{\alpha}\rfloor V + (e_{\alpha}\rfloor T^{\beta})\wedge H_{\beta} 
+ (e_{\alpha}\rfloor R^{\beta\gamma})\wedge H_{\beta\gamma},\label{Ea}\\
D\,{\mathcal E}_\alpha &\equiv (e_{\alpha}\rfloor T^{\beta})\wedge{\mathcal E}_\beta 
+ (e_{\alpha}\rfloor R^{\beta\gamma})\wedge\,{\mathcal C}_{\beta\gamma},\label{1st}
\end{align}
whereas the local Lorentz invariance results in the {\it Noether identity}
\begin{equation}
2D{\mathcal C}_{\alpha\beta} + \vartheta_\alpha\wedge{\mathcal E}_\beta 
- \vartheta_\beta\wedge{\mathcal E}_\alpha \equiv 0\,.\label{2nd}
\end{equation}
These relations are easily derived from (\ref{momC}), (\ref{conmomC}), and (\ref{Noe2}) by replacing $L$ with $V$ and dropping the dependence on $\psi^A$.  All the relations (\ref{Ea})-(\ref{2nd}) are {\it strong identities}; they are always valid independently of the field equations.

\subsection{The general field equations of Poincar\'e gauge gravity}

The field equations for the system of interacting matter and
gravitational fields are derived from the total Lagrangian
\begin{equation}
V(\vartheta^{\alpha}, T^{\alpha}, R^{\alpha\beta}) + 
L(\psi^A, D\psi^A, \vartheta^{\alpha}, T^{\alpha}, R^{\alpha\beta}).\label{Ltot}
\end{equation}
Independent variations of the total Lagrangian with respect to the coframe $\vartheta^\alpha$, the local Lorentz connection $\Gamma^{\alpha\beta}$, and the matter field $\psi^A$ yield the system \cite{vdHeyde:1976a,Hehl:1976kj,Hehl:EinsteinVolume,PRs}
\begin{align}
  DH_{\alpha}  - E_{\alpha} &= \Sigma_{\alpha}\,,\label{Peq1}\\
  DH_{\alpha\beta} - E_{\alpha\beta} &= {\frac 12}\,\tau_{\alpha\beta}
                                          \,,\label{Peq2}\\
  {\frac {\partial  L}{\partial\psi^A}}
  - (-1)^{p}D\,{\frac {\partial L}{\partial (D\psi^A)}} &= 0\,.\label{Pmat}
\end{align} 
From the dimension of the action we conclude that the Lagrangian 4-form has the same dimension $[L] = [\hbar]$, and one can easily find the dimension of the currents.  Taking into account that $[\vartheta^\alpha] = [\ell]$ and $[\Gamma^{\alpha\beta}] = [1]$ (dimensionless), using the definitions (\ref{sigC0}) and (\ref{spin0}) we derive the dimensions: $[\Sigma_{\alpha}] = [{\frac {\hbar} {\ell}}]$ = [momentum], and $[\tau_{\alpha\beta}] = [\hbar]$ = [spin].

By expanding the matter currents with respect to the basis of the 3-forms, we find the energy-momentum and the spin tensors, respectively, as
\begin{equation}
  \Sigma_\alpha = {\cal T}_\alpha{}^\mu\,\eta_\mu,\qquad
  \tau_{\alpha\beta} = {\cal S}_{\alpha\beta}{}^\mu\,\eta_\mu,\label{TS}
\end{equation}
with $\eta_\mu=\sqrt{-g}\,\epsilon_\mu$. Since $[\eta_\mu] = [\ell^3]$, we have the dimensions: $[{\cal T}_\alpha{}^\mu] = [{\frac {\hbar}{\ell^4}}]$ = [momentum/volume], and $[{\cal S}_{\alpha\beta}{}^\mu] = [{\frac {\hbar}{\ell^3}}]$ = [spin/volume].

\subsection{Einstein-Cartan theory}

The simplest Poincar\'e gauge theory is the Einstein-Cartan theory of gravity (EC), which results from choosing the {\it curvature scalar} of the Riemann-Cartan space as gravitational Hilbert-Einstein Lagrangian ($\eta_{\alpha\beta}=\,^*(\vartheta_\alpha\wedge\vartheta_\beta)$, see Appendix 1)
\begin{equation}
V_{\text{HE}} = {\frac {1}{2\kappa}}\eta_{\alpha\beta}\wedge R^{\alpha\beta}.\label{LHE}
\end{equation}
Here, $\kappa = {\frac {8\pi G}{c^3}}$ is Einstein's gravitational constant with the dimension of $[\kappa] = $\,s\,kg$^{-1}$. $G = 6.67\times 10^{-11}$ m$^3$\,kg$^{-1}$\,s$^{-2}$ is Newton's gravitational constant. The speed of light $c = 2.9\times 10^8$ m/s. Consistency check for the dimension: $[{\frac {1}{\kappa}}] =\,$kg s$^{-1}$ = $[{\frac mt}]$. Since $[R^{\alpha\beta} ] = 1$ and $[\eta_{\alpha\beta}] = \ell^2$, we have $[{\frac {1}{2\kappa}}\eta_{\alpha\beta} \wedge R^{\alpha\beta}] = [{\frac {m\ell^2}{t}}] = [\hbar]$.

For the Lagrangian (\ref{LHE}) we find from (\ref{HH}), (\ref{EE}) and (\ref{Ea}):
\begin{equation}
H_\alpha = 0,\quad H_{\alpha\beta} = {\frac {1}{2\kappa}}\eta_{\alpha\beta},\quad
E_\alpha = -\,{\frac {1}{2\kappa}}\eta_{\alpha\beta\gamma}\wedge R^{\beta\gamma},
\quad E_{\alpha\beta} = 0.\label{HHEE}
\end{equation}
As a result, 
\begin{equation}\label{EC}
{\cal E}_\alpha = {\frac {1}{2\kappa}}\eta_{\alpha\beta\gamma}\wedge R^{\beta\gamma},\qquad 
{\cal C}_{\alpha\beta} = {\frac {1}{2\kappa}}\eta_{\alpha\beta\gamma}\wedge T^{\gamma},
\end{equation}
and hence the Einstein-Cartan gravitational field equations read
\begin{equation}
 {\frac {1}{2}}\eta_{\alpha\beta\gamma}\wedge R^{\beta\gamma} = \kappa\,\Sigma_\alpha,\qquad
\eta_{\alpha\beta\gamma}\wedge T^{\gamma} = \kappa\,\tau_{\alpha\beta}.\label{ECeq}
\end{equation}

Substituting $R^{\alpha\beta} = {\frac 12}R_{\mu\nu}{}^{\alpha\beta}\,\vartheta^\mu\wedge\vartheta^\nu$ and
$T^{\alpha} = {\frac 12}T_{\mu\nu}{}^\alpha\,\vartheta^\mu\wedge\vartheta^\nu$ into the left-hand side of (\ref{ECeq}) and using (\ref{TS}), we find the Einstein-Cartan field equations in components,
\begin{align}\label{EC1}
  {\rm Ric}_\alpha{}^\beta - {\frac 12}\delta_\alpha^\beta\,{\rm Ric}_\gamma{}^\gamma
  &= \kappa\,{\cal T}_\alpha{}^\beta,\\
  T_{\alpha\beta}{}^\gamma - \delta_\alpha^\gamma T_{\mu\beta}{}^\mu + \delta_\beta^\gamma T_{\mu\alpha}{}^\mu
  &= \kappa\,{\cal S}_{\alpha\beta}{}^\gamma.\label{EC2}
\end{align} 
From the curvature 2-form we derive the Ricci 1-form ${\rm Ric}_\alpha = e_\beta\rfloor R_\alpha{}^\beta = {\rm Ric}_{i\alpha} dx^i$, components of which constitute the Ricci tensor ${\rm Ric}_{i\alpha} = e^j_\beta R_{ji\alpha}{}^\beta$. The curvature scalar is defined as usual by $R = e^\alpha\rfloor{\rm Ric}_\alpha = e_\alpha\rfloor e_\beta\rfloor R^{\alpha\beta} = e^i_\beta e^j_\alpha R_{ij}{}^{\alpha\beta}$.

One can ``minimally'' extend the EC theory by modifying the Hilbert-Einstein Lagrangian (\ref{LHE}) with a simplest possible parity-odd term. Such a generalization was proposed by Hojman, Mukku \& Sayed \cite{Hojman:1980kv} (and is known as the Holst Lagrangian \cite{Holst} in some literature) as
\begin{equation}
V_{\text{HMS}} = {\frac {1}{2\kappa}}\left(\eta_{\alpha\beta} + \overline{a}{}_0
\vartheta_\alpha\wedge\vartheta_\beta\right)\wedge R^{\alpha\beta}.\label{LH}
\end{equation}
Here $\overline{a}{}_0 = {\frac 1\xi}$, and the dimensionless constant $\xi$ is often
called the Barbero-Immirzi parameter.

For the Lagrangian (\ref{LH})---due to (\ref{HH}), (\ref{EE}), and (\ref{Ea})---we find $H_\alpha = 0$ and $E_{\alpha\beta} = 0$ as in (\ref{HHEE}); however, now
\begin{equation}
H_{\alpha\beta} = {\frac {1}{2\kappa}}\left(\eta_{\alpha\beta} + \overline{a}{}_0
\vartheta_\alpha\wedge\vartheta_\beta\right),\quad
E_\alpha = -\,{\frac {1}{2\kappa}}\left(\eta_{\alpha\beta\gamma}\wedge R^{\beta\gamma} + 
2\overline{a}{}_0R_{\alpha\beta}\wedge\vartheta^\beta\right).\label{HE2}
\end{equation}
As a result, ${\cal E}_\alpha = -\,E_\alpha$, and
\begin{equation}\label{CH}
{\cal C}_{\alpha\beta} = DH_{\alpha\beta} = {\frac {1}{2\kappa}}\left[
\eta_{\alpha\beta\gamma}\wedge T^{\gamma} + \overline{a}{}_0\left(T_\alpha\wedge\vartheta_\beta 
- T_\beta\wedge\vartheta_\alpha\right)\right].
\end{equation}
The Einstein-Cartan field equations (\ref{ECeq}) are then replaced by
\begin{align}\label{ECH1}
{\frac 12}\eta_{\alpha\beta\gamma}\wedge R^{\beta\gamma} + \overline{a}{}_0R_{\alpha\beta}
\wedge\vartheta^\beta &= \kappa\,\Sigma_\alpha,\\
\eta_{\alpha\beta\gamma}\wedge T^{\gamma} + \overline{a}{}_0\left(T_\alpha\wedge\vartheta_\beta
- T_\beta\wedge\vartheta_\alpha\right) &=\kappa\,\tau_{\alpha\beta}.\label{ECH2}
\end{align}

The Einstein-Cartan theory is a viable gravitational theory that deviates from GR only at very high matter density, see \cite{Boos:2017} and the references given there. For matter without spin, ${\cal S}_{\a\b}{}^{\g} = 0$, the torsion vanishes because of the second field equation (\ref{EC2}) ,and the Einstein-Cartan theory coincides with GR. The same applies to the parity-odd HMS model (\ref{LH}).

\subsection{Quadratic Poincar\'e gauge gravity models}

It is a unique feature of the gauge approach to gravity that the field equations (\ref{EC1}) and (\ref{EC2}) are {\it algebraic} in the field strength rather than first order partial differential equations as in any other gauge theory of an internal group $U(1)$, $SU(2)$, $SU(3)$,...\ . This is due to the fact that, because of the existence of the metric $g_{\a\b}$ of spacetime and of the frame $e_i{}^\a$, with $e_i{}^\a\,e^{i}{}_{\b}=\d^\a_\b$, the Lorentz field strength $R_{\a\b}{}^{\g\d}=e^i{}_\a e^j{}_\b\,R_{ij}{}^{\g\d}$ can be contracted to a scalar, namely to the curvature scalar.

This is impossible in Yang-Mills theories since the internal group indices cannot be related to the spacetime indices. Thus, the simplest Lagrangian in Yang-Mills theory is {\it quadratic} in the field strength yielding first-order differential equations as field equations. Accordingly, the gauge doctrine would suggest for the gauge theory of the Poincar\'e group a general Lagrangian 4-form that contains all possible quadratic invariants of the torsion and the curvature:
\begin{align}
V =& \,{\frac {1}{2\kappa}}\Big\{\Big(a_0\eta_{\alpha\beta} + \overline{a}_0
\vartheta_\alpha\wedge\vartheta_\beta\Big)\wedge R^{\alpha\beta} - 2\lambda_0\eta \nonumber\\
& -\,T^\alpha\wedge\sum_{I=1}^3
\left[a_I\,{}^*({}^{(I)}T_\alpha) + \overline{a}_I\,{}^{(I)}T_\alpha\right]\Big\}\nonumber\\
& - \,{\frac 1{2\rho}}R^{\alpha\beta}\wedge\sum_{I=1}^6 \left[b_I\,{}^*({}^{(I)}\!R_{\alpha\beta}) 
+ \overline{b}_I\,{}^{(I)}\!R_{\alpha\beta}\right].\label{LRT}
\end{align}
The irreducible parts of the torsion and the curvature are defined in Appendix~2.  The Lagrangian (\ref{LRT}) has a clear structure: the first line is {\it linear in the curvature,} the second line collects the terms {\it quadratic in the torsion}, while the third line contains the invariants {\it quadratic in the curvature}. Furthermore, each line is composed of a parity even piece (first term on each line), and a parity odd part (last term on each line). A special case $a_0 = 0$ and $\overline{a}_0 = 0$ describes the purely quadratic model without the Hilbert-Einstein linear term in the Lagrangian. To recover the Einstein-Cartan model, one puts $a_0 = 1$ and $\overline{a}_0 = 0$.

Besides that, the general PG model contains a set of the coupling constants which determine the structure of quadratic part of the Lagrangian: $\rho$, $a_1, a_2, a_3$ and $\overline{a}_1, \overline{a}_2, \overline{a}_3$, $b_1, \cdots, b_6$ and $\overline{b}_1, \cdots, \overline{b}_6$. The overbar denotes the constants responsible for the parity odd interaction. We have the dimension $[{\frac 1\rho}] = [\hbar]$, whereas $a_I$, $\overline{a}_I$, $b_I$ and $\overline{b}_I$ are dimensionless. Moreover, not all of these constants are independent: we take $\overline{a}_2 = \overline{a}_3$, $\overline{b}_2 = \overline{b}_4$ and $\overline{b}_3 = \overline{b}_6$ because some of terms in (\ref{LRT}) are the same,
\begin{align}\label{T23}
T^\alpha\wedge{}^{(2)}T_\alpha = T^\alpha\wedge{}^{(3)}T_\alpha = {}^{(2)}T^\alpha\wedge{}^{(3)}T_\alpha,\\
R^{\alpha\beta}\wedge{}^{(2)}\!R_{\alpha\beta} = R^{\alpha\beta}\wedge{}^{(4)}\!R_{\alpha\beta} 
= {}^{(2)}\!R^{\alpha\beta}\wedge{}^{(4)}\!R_{\alpha\beta},\label{R24} \\ 
R^{\alpha\beta}\wedge{}^{(3)}\!R_{\alpha\beta} = R^{\alpha\beta}\wedge{}^{(6)}\!R_{\alpha\beta} 
= {}^{(3)}\!R^{\alpha\beta}\wedge{}^{(6)}R_{\alpha\beta}.\label{R36}
\end{align}

For the Lagrangian (\ref{LRT}), by means of (\ref{HH})-(\ref{EE}), we
can derive the gravitational field momenta
\begin{equation}
H_\alpha = -\,{\frac 1{\kappa}}\,h_\alpha\,,\qquad H_{\alpha\beta} = {\frac {1}{2\kappa}}
\left(a_0\,\eta_{\alpha\beta} + \overline{a}_0\vartheta_\alpha\wedge\vartheta_\beta\right)
- {\frac 1\rho}\,h_{\alpha\beta},\label{HabRT}
\end{equation}
and, furthermore, the canonical energy-momentum and spin currents of the gravitational field as
\begin{align}
E_\alpha &= -\,{\frac {1}{\kappa}}\Big({\frac
           {a_0}2}\,\eta_{\alpha\beta\gamma}\wedge R^{\beta\gamma}
+ \overline{a}_0\,R_{\alpha\beta}\wedge\vartheta^\beta 
- \lambda_0\eta_\alpha + q^{(T)}_\alpha\Big)
-\, {\frac 1\rho}\,q^{(R)}_\alpha,\label{EaRT}\\
E_{\alpha\beta} &= {\frac 12}\left(H_\alpha\wedge\vartheta_\beta 
- H_\beta\wedge\vartheta_\alpha\right).\label{EabRT}
\end{align}
For convenience, we introduced here the 2-forms which are linear
functions of the torsion and the curvature, respectively, by
\begin{equation}
h_\alpha = \sum_{I=1}^3\left[a_I\,{}^*({}^{(I)}T_\alpha) 
+ \overline{a}_I\,{}^{(I)}T_\alpha\right],\quad 
h_{\alpha\beta} = \sum_{I=1}^6\left[b_I\,{}^*({}^{(I)}\!R_{\alpha\beta}) 
+ \overline{b}_I\,{}^{(I)}\!R_{\alpha\beta}\right],\label{hR}
\end{equation}
and 3-forms quadratic in the torsion and in the curvature, respectively:
\begin{align}
  q^{(T)}_\alpha &= {\frac 12}\left[(e_\alpha\rfloor T^\beta)\wedge h_\beta
   - T^\beta\wedge e_\alpha\rfloor h_\beta\right],\label{qa}\\
  q^{(R)}_\alpha &= {\frac 12}\left[(e_\alpha\rfloor R^{\beta\gamma})\wedge h_{\beta\gamma} 
   - R^{\beta\gamma}\wedge e_\alpha\rfloor h_{\beta\gamma} \right].\label{qaR}
\end{align}
By construction, the first 2-form in (\ref{hR}) has the dimension of a length, $[h_\alpha] = [\ell]$, whereas the second one is obviously dimensionless, $[h_{\alpha\beta}] = 1$. Similarly, we find for (\ref{qa}) the dimension of length $[q^{(T)}_\alpha] = [\ell]$, and the dimension of the inverse length, $[q^{(R)}_\alpha] = [1/\ell]$ for (\ref{qaR}).

The resulting Poincar\'e gravity field equations (\ref{dVt}) and (\ref{dVG}) then read:
\begin{align}
{\frac {a_0}2}\eta_{\alpha\beta\gamma}\wedge R^{\beta\gamma} + \overline{a}_0R_{\alpha\beta}
\wedge\vartheta^\beta - \lambda_0\eta_\alpha & \nonumber\\ 
+ \,q^{(T)}_\alpha + \ell_\rho^2\,q^{(R)}_\alpha - Dh_\alpha &= \kappa\,\Sigma_\alpha,\label{ERT1}\\
a_0\,\eta_{\alpha\beta\gamma}\wedge T^{\gamma} + \overline{a}_0\left(T_\alpha\wedge\vartheta_\beta
- T_\beta\wedge\vartheta_\alpha \right) & \nonumber\\ 
+ \,h_\alpha\wedge\vartheta_\beta - h_\beta\wedge\vartheta_\alpha 
- 2\ell_\rho^2\,Dh_{\alpha\beta} &= \kappa\,\tau_{\alpha\beta}.\label{ERT2}
\end{align}
The contribution of the curvature square terms in the Lagrangian (\ref{LRT}) to the gravitational field dynamics in the equations (\ref{ERT1}) and (\ref{ERT2}) is characterized by the new coupling parameter with the dimension of the area (recall that $[{\frac 1\rho}] = [\hbar]$):
\begin{equation}
\ell_\rho^2 = {\frac {\kappa}{\rho}}.\label{lr}
\end{equation}
The parity-odd sector in PG gravity has been recently analyzed in \cite{Diakonov,Baekler1,Baekler2,Chen,Ho2,Ho3,Ho4}, with a particular attention to the cosmological issues. A major progress was made in this domain with the computation of the particle spectrum of general quadratic PG models by Karananas \cite{Karananas} and Blagojevi\'c and Cvetkovi\'c \cite{Blagojevic:2018dpz}.

\begin{figure}
\vspace{10cm}
\centering
  \includegraphics[width=0.7\textwidth]{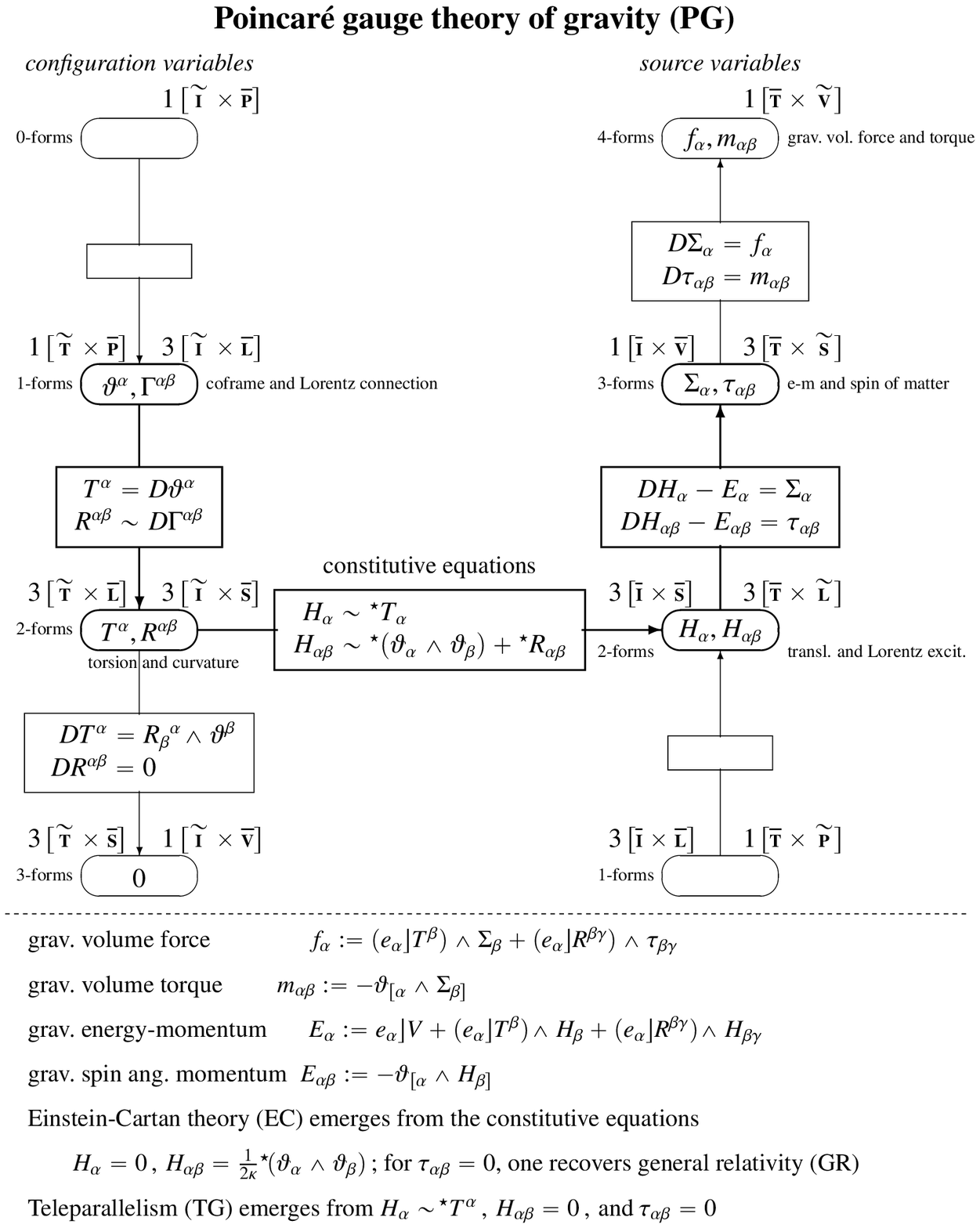}
  \caption{Tonti diagram for Poincar\'e gauge gravity theory.}\label{TontiPG}
\end{figure}

\subsection{Tonti-diagram of quadratic Poincar\'e gauge gravity}

The structure of Poincar\'e gauge gravity can be visualized by means of the Tonti diagram in Fig.~\ref{TontiPG}.  It is constructed as a direct generalization of the Tonti diagram for TG in Fig.~\ref{TontiTG}. In TG we started with 4 translation potentials $\vt^\a$, in PG we start instead with $4$ translation and $6$ Lorentz 1-form potentials as {\it configuration} variables: $\{\vt^\a,\Gamma^{\a\b}=-\Gamma^{\b\a}\}$. Besides the translation group $T(4)$, now the Lorentz group $SO(1,3)$ is gauged, too. On the side of the {\it source} variables, besides the energy-momentum current $\Sigma_\a$, we have the spin current $\tau_{\a\b}=-\tau_{\b\a}$.

Since the Lorentz group depends on the metric, the present Tonti diagram does not contain premetric structures. However, if we generalize the Lorentz connection $\Gamma^{\a\b}$ to a linear one, $\Gamma_\a{}^\b$, that is, if we substitute the $SO(1,3)$ by the linear group $GL(4,R)$, then the configuration and source variables are all premetric again, similar as in the Tonti diagram for TG. Note that $\Gamma^{\a\b}:=g^{\a\g}\,\Gamma_\g{}^\b$. The affine generalization of PG has been discussed as metric-affine gravity in detail in \cite{PRs}, for example.

As we recognize from the Tonti diagram for PG, the {\it linear} constitutive equations correspond to quadratic Lagrangians, and the Einstein-Cartan theory and Einstein's general relativity are simple subcases of this scheme.

\section{Discussion and outlook}

A gauge theory is a heuristic scheme within the Lagrange formalism in the Minkowski space of special relativity for the purpose of deriving a new interaction from a conserved current and the attached rigid symmetry group. This new `gauge' interaction is induced by demanding that the rigid symmetry should be extended to a locally valid symmetry. We demonstrate that, when applied to gravity theory, this heuristic approach should be based on the conserved energy-momentum current and the associated translational symmetry, consistently supplemented by the local Lorentz symmetry.

We postpone a detailed review of the physical contents of the Poincar\'e gauge gravity theory to a different publication. Earlier, this subject was intensively studied and the relevant results can be found in \cite{vdHeyde:1976a,Hehl:1976kj,Hehl:EinsteinVolume,Shapiro}.

At present time, we are again witnessing a considerable growth of the interest to the gauge gravitational issues. Among other directions of research, the search and analysis of {\it exact solutions} of the field equations is essential for improvement of understanding of the nature of gravitational interaction; see \cite{Cembranos1,Cembranos2,Heinicke:2015iva,Obukhov:2019} for relevant discussions. Also the topological invariants related to torsion have been reinvestigated \cite{Nieh:2017,Nieh:2018rlg}.

It is important to recall Einstein's view \cite{Einstein:1921GE} that ``...the question whether this continuum has a Euclidean, Riemannian, or any other structure is a question of physics proper which must be answered by experience, and not a question of a convention to be chosen on grounds of mere expediency.'' Accordingly, when aiming at an experimental probing of the geometrical structure of spacetime (in particular, searching for possible deviations beyond the Riemannian geometry), one should study how do test particles move under the influence of the gravitational field in the framework of the gauge gravity theory. Remarkably, the {\it propagation equations} should not be postulated in an ad hoc way; they are the consequence of the conservation laws. Importantly, the analysis of the equations of motion, reveals that the torsion, in the context of PG, is related only to the {\it elementary particle spin} and under no circumstances to the orbital angular momentum of test particles \cite{vdHeyde:1976b,Yasskin,Hehl:2013,Obukhov:2015eqa}.

Recent advances in {\it cosmology} have drawn renewed attention to the framework of Poincar\'e gauge gravity. In this sense, the early results of Trautman \cite{Trautman:1973}, and Minkevich \cite{Mink1,Mink2}, which predict possible avertion of singularity in the early universe and possible modifications of the late stage of cosmological evolution \cite{Magu,Pop,Dirk,Zhang:2019} are currently revisited and extended in the most recent works of Barrow et al.\ \cite{Kranas:2018,Barrow:2019} and of Nikiforova, Randjbar-Daemi, Rubakov and Damour \cite{Niki:2009,Niki:2017,Damour:2018,Damour:2019}.

Last but not least we should mention the progress in the study of the Hamiltonian approach to PG by Struckmeier et al.\ \cite{Struckmeier:2015,Struckmeier:2017,Struckmeier:2018hjo} and, most remarkably, by Blagojevi\'c and Cvetkovi\'c \cite{Blagojevic:2019a,Blagojevic:2019b}. Toller \cite{Toller:2017} has proposed a highly interesting generalization of the PG framework by taking the symplectic group $Sp(4,R)$ as gauge group, which is locally isomorphic to the anti-de Sitter group $SO(2,3)$.

Finally, let us have a look back at the Tonti diagram for TG, see Fig.~\ref{TontiTG}. The constitutive equation is conventionally assumed to be a linear relation. Only then we can recover GR by taking a suitable ansatz. Mashhoon, already since the early 1990s, followed up the idea that the locality principle in special relativity---the clock hypothesis is one particular example---is washed out at extremely high translational and rotational accelerations. This is expected to happen long before quantum effects set in. A detailed discussion can be found in Mashhoon's book \cite{Mashhoon:NLG}. This nonlocality should also have consequences for gravitational theory. The linear constitutive equation in Fig.~\ref{TontiTG} should then become a {\it nonlocal} relation, as was proposed by Mashhoon and one of us \cite{Hehl:2008eu,Hehl:2009es} and has been discussed in detail in \cite{Mashhoon:NLG}. A certain simplification of the nonlocal ansatz has been proposed by Puetzfeld et al.\ \cite{Puetzfeld:2019}. The nonlocal idea, as applied to TG, has even be generalized to the complete quadratic PG framework, see \cite[Appendix]{Blome:2010xn}. But a detailed discussion is still missing.

\section*{Appendix 1}
\addcontentsline{toc}{section}{Appendix 1}

Our basic notation and conventions are consistent with \cite{Birkbook,PRs}. In particular, Greek indices $\alpha, \beta, \dots = 0, \dots, 3$, denote the anholonomic components (for example, of a coframe $\vartheta^\alpha$), while the Latin indices $i,j,\dots =0,\dots, 3$, label the holonomic components (e.g., $dx^i$). Spatial components are numbered by Latin indices from the beginning of the alphabet $a, b, \dots = 1, 2, 3$. The anholonomic vector frame basis $e_\alpha$ is dual to the coframe basis in the sense that $e_\alpha\rfloor\vartheta^\beta = \delta_\alpha^\beta$, where $\rfloor$ denotes the interior product. The volume 4-form is denoted by $\eta$, and the $\eta$-basis in the space of exterior forms is constructed with the help of the interior products as $\eta_{\alpha_1 \dots\alpha_p}:= e_{\alpha_p}\rfloor\dots e_{\alpha_1}\rfloor\eta$, $p=1,\dots,4$. They are related to the $\vartheta$-basis via the Hodge dual operator $^*$, for example, $\eta_{\alpha\beta} = {}^*\!\left(\vartheta_\alpha\wedge\vartheta_\beta\right)$. The Minkowski metric is $g_{\alpha\beta} = {\rm diag}(c^2,-1,-1,-1)$. All the objects related to the parity-odd sector (coupling constants, irreducible pieces of the curvature, etc.) are marked by an overline, to distinguish them from the corresponding parity-even objects.

\section*{Appendix 2}
\addcontentsline{toc}{section}{Appendix 2}

The torsion 2-form can be decomposed into the three irreducible pieces,
$T^{\alpha}={}^{(1)}T^{\alpha} + {}^{(2)}T^{\alpha} + {}^{(3)}T^{\alpha}$, where
\begin{align}
{}^{(2)}T^{\alpha} &= {\frac 13}\vartheta^{\alpha}\wedge T,\qquad 
{}^{(3)}T^{\alpha} = {\frac 13}e^\alpha\rfloor{}^\ast \overline{T},\label{iT23}\\
{}^{(1)}T^{\alpha} &= T^{\alpha}-{}^{(2)}T^{\alpha} - {}^{(3)}T^{\alpha}.\label{iT1}
\end{align}
Here the 1-forms of the torsion trace and axial trace are introduced:
\begin{equation}
T := e_\nu\rfloor T^\nu,\qquad \overline{T} := {}^*(T^{\nu}\wedge\vartheta_{\nu}).\label{traces1}
\end{equation}

For the irreducible pieces of the dual torsion ${}^*T^{\alpha} = {}^{(1)}({}^*T^{\alpha}) + {}^{(2)}({}^*T^{\alpha}) + {}^{(3)}({}^*T^{\alpha})$, we have the properties 
\begin{equation}
{}^{(1)}({}^*T^\alpha)={}^*({}^{(1)}T^\alpha),\quad
{}^{(2)}({}^*T^\alpha)={}^*({}^{(3)}T^\alpha),\quad
{}^{(3)}({}^*T^\alpha)={}^*({}^{(2)}T^\alpha).\label{dTdual}
\end{equation}

The Riemann-Cartan curvature 2-form is decomposed $R^{\alpha\beta} = \sum_{I=1}^6\,{}^{(I)}\!R^{\alpha\beta}$ into the 6 irreducible parts 
\begin{eqnarray}
&{}^{(2)}\!R^{\alpha\beta} = -\,{}^*(\vartheta^{[\alpha}\wedge\overline{\Psi}{}^{\beta]}),\qquad
{}^{(4)}\!R^{\alpha\beta} = -\,\vartheta^{[\alpha}\wedge\Psi^{\beta]},& \label{curv24}\\
&{}^{(3)}\!R^{\alpha\beta} = -\,{\frac 1{12}}\,\overline{X}\,{}^*\!(\vartheta^\alpha\wedge
\vartheta^\beta),\qquad {}^{(6)}\!R^{\alpha\beta}  = -\,{\frac 1{12}}\,X\,\vartheta^\alpha\wedge
\vartheta^\beta,& \label{curv36}\\
&{}^{(5)}\!R^{\alpha\beta} = -\,{\frac 12}\vartheta^{[\alpha}\wedge e^{\beta]}
\rfloor(\vartheta^\gamma\wedge X_\gamma),& \label{curv5}\\
&{}^{(1)}\!R^{\alpha\beta} = R^{\alpha\beta} - \sum\limits_{I=2}^6\,{}^{(I)}\!R^{\alpha\beta},& \label{curv1}
\end{eqnarray}
where 
\begin{equation}
X^\alpha := e_\beta\rfloor R^{\alpha\beta},\quad X := e_\alpha\rfloor X^\alpha,
\quad \overline{X}{}^\alpha := {}^*(R^{\beta\alpha}\wedge\vartheta_\beta),\quad 
\overline{X} := e_\alpha\rfloor \overline{X}{}^\alpha,\label{WX}
\end{equation}
and 
\begin{eqnarray}
&\Psi_\alpha := X_\alpha - {\frac 14}\,\vartheta_\alpha\,X - {\frac 12}
\,e_\alpha\rfloor (\vartheta^\beta\wedge X_\beta),&\label{Psia}\\
&\overline{\Psi}_\alpha := \overline{X}_\alpha - {\frac 14}\,\vartheta_\alpha
\,\overline{X} - {\frac 12}\,e_\alpha\rfloor (\vartheta^\beta\wedge \overline{X}_\beta).&\label{Phia}
\end{eqnarray}
The 1-forms $X^\alpha$ and $\overline{X}{}^\alpha$ are not completely independent: $\vartheta_\alpha\wedge X^\alpha = {}^*(\vartheta_\alpha\wedge \overline{X}{}^\alpha)$.

\subsection*{Acknowledgments}

This article is based on a lecture given at the Weyl Conference in Bad Honnef in 2018. FWH is very grateful to Claus Kiefer, Silvia De Bianchi, and to the Wilhelm and Else Heraeus Foundation for the invitation. The material of the paper is partly overlapping with lectures presented in Tartu in 2017 at the conference ``Geometric Foundations of Gravity'' organized by Manuel Hohmann, Christian Pfeifer, Laur J\"arv and Martin Kr\v{s}\v{s}\'ak. We thank Yakov Itin (Jerusalem) and Jens Boos (Edmonton, Alberta) for many fruitful discussions. We are specially thankful to Prof.\ Enzo Tonti (Trieste) for his help in setting up the two Tonti diagrams. We are most grateful to Claus Kiefer for numerous suggestions for improving our article.

\end{document}